\documentclass{aa} 

\usepackage{graphicx}	
\usepackage{changepage}

\usepackage{bm}
\usepackage{txfonts}
\usepackage{amsmath}	
\usepackage{amssymb}	
\usepackage{amsfonts}	
\usepackage{tabularx}
\usepackage{hyperref}
\usepackage{graphicx}	
\usepackage{siunitx} 
\usepackage{color}

\def\lsim{\mathrel{\rlap{\lower 3pt \hbox{$\sim$}} \raise 2.0pt \hbox{$<$}}}
\def\gsim{\mathrel{\rlap{\lower 3pt \hbox{$\sim$}} \raise 2.0pt \hbox{$>$}}}

\hypersetup{draft}
\begin{document}

\title{Chemical analysis of prestellar cores in Ophiuchus yields short timescales and rapid collapse}

\subtitle{}

\author{
	Stefano Bovino\inst{1}\thanks{E-mail: stefanobovino@udec.cl}
	\and
	Alessandro Lupi\inst{2}
	\and
	Andrea Giannetti\inst{3}
	\and
	Giovanni Sabatini\inst{1,3,4}
	\and 
	Dominik R. G. Schleicher\inst{1}
	\and
	Friedrich Wyrowski\inst{5}
	\and
	Karl M. Menten\inst{5}
	}

\institute{
    Departamento de Astronom\'ia, Facultad Ciencias F\'isicas y Matem\'aticas, Universidad de Concepci\'on\\
    Av. Esteban Iturra s/n Barrio Universitario, Casilla 160, Concepci\'on, Chile
    \and
	Dipartimento di Fisica G.~Occhialini", Universit\`{a} degli Studi di Milano-Bicocca, Piazza della Scienza 3, IT-20126 Milano, Italy
	\and
	INAF - Istituto di Radioastronomia - Italian node of the ALMA Regional Centre (ARC), via Gobetti 101, I-40129 Bologna, Italy
	\and
	Dipartimento di Fisica e Astronomia ``Augusto Righi'', Universit\'a degli Studi di Bologna, via Gobetti 93/2, I-40129 Bologna, Italy
	\and
	Max-Planck-Institut f\"ur Radioastronomie, Auf dem H\"ugel, 69, 53121, Bonn Germany
}

\date{Accepted XXX. Received YYY; in original form ZZZ}

\abstract{
Sun-like stars form from the contraction of cold and dense interstellar clouds. 
How the collapse proceeds and what are the main physical processes driving it, however, is still under debate and a final consensus on the timescale of the process has not been reached. Does this contraction proceed slowly, sustained by strong magnetic fields and ambipolar diffusion, or is it driven by fast collapse with gravity dominating the entire process? One way to answer this question is to measure the age of prestellar cores through statistical methods based on observations or via reliable chemical chronometers, which should better reflect the physical conditions of the cores.
Here we report APEX observations of ortho-H$_2$D$^+$ and para-D$_2$H$^+$ for six cores in the Ophiuchus complex and combine them with detailed three-dimensional magneto-hydrodynamical simulations including chemistry, providing a range of ages for the observed cores of  up to 200 kyr. The outcome of our simulations and subsequent analysis provides a good matching with the observational results in terms of physical (core masses and volume densities) and dynamical parameters such as the Mach number and the virial parameter.
We show that models of fast collapse successfully reproduce the observed range of chemical abundance ratios as the timescales to reach the observed stages is comparable to the dynamical time of the cores (i.e. the free-fall time) and much shorter than the ambipolar diffusion time, measured from the electron fraction in the simulations. To confirm that this ratio can be used to distinguish between different star-formation scenarios a larger (statistically relevant) sample of star-forming cores should be explored.}


\maketitle


\section{Introduction}
Star formation is a longstanding problem in astrophysics  \citep{Bergin2007}, and despite the observational and theoretical progress made over the last decades, there are still some fundamental questions that remain open.  One of these relates to the timescale to go from the prestellar stage to the formation of a protostellar object. The latter is affected by the interplay of the physical processes that drive star-formation during the gravitational collapse of the gas within the prestellar cores. Strong magnetic fields and ambipolar diffusion \citep{Shu1987,Mouschovias1991}, tend to slow-down the collapse, which occurs on timescales longer than the typical dynamical time (i.e. the free-fall time), while in fast collapse theories the process is dominated by gravity and the results are timescales comparable to the free-fall time \citep{Enrique2005,Hartmann2012}. 

Measuring timescales and assessing the age of prestellar cores is challenging. One way is to employ statistical methods based on observations 
\citep{Lee1999}. A viable alternative is to use chemical clocks, i.e. tracers which show strong monotonic behaviour with time  \citep{Pagani2013,Brunken2014}, and better reflect the physical conditions of the cores. This is particularly true in star-forming regions, as the star formation process itself goes through strong changes in density and temperature, which in turn affect the chemical evolution of specific tracers, sensitive to these physical quantities.

\begin{table*}
\centering
\caption{Observational properties of the sources in Ophiuchus. We report the dust temperature $T_\mathrm{dust}$, the core size $R_\mathrm{eff}$, the core mass $M_\mathrm{core}$, the Jeans mass ($M_\mathrm{J}$), the logarithm of the H$_2$ column density, and the total number density $n_\mathrm{core}$ = $M_\mathrm{core}/(V_\mathrm{core}\mu m_\mathrm{H})$, with $V_\mathrm{core}$ the core volume. For more details on how these quantities are calculated we refer to the formulae reported in Appendix A.}\label{table:prop}
\vspace{0.2mm}
\setlength{\tabcolsep}{2pt}
\begin{tabular*}{\textwidth}{@{\extracolsep{\fill} }ccccccccc}
\hline
\hline

\#  &   RA$_{2000}$ & DEC$_{2000}$ & $T_{\rm dust}$ & $R_{\rm eff}$ & $M_{\rm core}$ & $M_{\rm J}$ &  log$_{10}$[N(H$_2$)] & log$_{10}$($n_\mathrm{core}$)  \\
core & [hh:mm:ss] & [dd:mm:ss] & [K] & [AU] & [M$_\odot$] & [M$_\odot$] & [cm$^{-2}]$ & [cm$^{-3}$]\\
\hline
1 & 16:31:57.637 & -24:57:35.70 & 12.00 & 4000 & 0.75 &  0.55 &22.48 & 5.67 \\
2 & 16:31:39.935 & -24:49:50.25 & 11.75 & 4890 & 2.09 &  0.43 &22.73 & 5.86 \\
3 &	16:27:33.241 & -24:26:24.09 & 10.00 & 3530 & 4.17 &  0.15 &22.97 & 6.58 \\
4 &	16:27:12.765 & -24:29:40.25 & 11.70 & 3775 & 1.41 &  0.35 &22.82 & 6.03 \\
5 &	16:27:15.277 & -24:30:30.06 & 12.80 & 3795 & 1.13 &  0.46 &22.83 & 5.91 \\
6 &	16:27:19.991 & -24:27:17.64 & 10.45 & 3010 & 1.53 &  0.21 &22.68 & 6.35 \\

\hline
\hline
\end{tabular*}
\end{table*}

The ideal chemical clock to trace the star formation process would be the ortho-to-para ratio of the most abundant molecule in these regions, molecular hydrogen, which is expected to decrease over time during cloud  contraction \citep{Pagani2009_2}, even if a decrease of this quantity should be already in place at the large scales of molecular cloud formation. The impossibility to access H$_2$ rotational transitions at the low temperatures of dense cores (\mbox{$T < 20$ K}) requires to employ a proxy, which manifests similar evolutionary features as H$_2$. 
The specific conditions of temperature and density (\mbox{$n(\mathrm{H_2}) >$ 10$^4$ cm$^{-3}$}) make dense cores the ideal places to boost 
the deuterium enrichment of key molecules, the so-called deuterium fractionation process \citep{Caselli2002P&SS,Ceccarelli2014}, which has been extensively employed as chemical clock and it is known to be strongly affected by changes in the ortho-to-para H$_2$ ratio.
Previous works focused on the ortho-to-para H$_2$D$^+$ ratio \citep{Brunken2014} and  N$_2$D$^+$/N$_2$H$^+$
 \citep{Caselli2002P&SS,Pagani2013}. However, the former one suffers the challenges of accessing the THz domain to observe para-H$_2$D$^+$, and is also expected to be converted in D$_2$H$^+$ over time, whereas the latter one does not necessarily trace the very cold and early stages \citep{Pillai2012,Giannetti2019}. N$_2$D$^+$ is also subject to depletion \citep{Pagani2007,Redaelli2019,Liu2020} making the  estimates of timescales less accurate.  
 
In addition, it is very difficult to generalise the effectiveness of the aforementioned clocks with a statistically relevant sample, due to the high level of THz continuum needed to detect the para-H$_2$D$^+$ line in absorption. Sources in very early stages of evolution simply do not satisfy this requirement. 
 
\citet{Vastel2004,Vastel2012}, \citet{Parise2011}, and \citet{Pagani2013} have also explored the ratio between ortho-H$_2$D$^+$ and para-D$_2$H$^+$, finding values around unity. The main advantage, compared to the para-H$_2$D$^+$, lies on the fact that both tracers are observable from the ground in emission, without the necessity of a strong background source \citep{Brunken2014}. However, in these works the possibility to use this ratio as a chemical clock was not convincingly established. In particular, \citet{Pagani2013}, based on low-dimensional models, claimed that the non monotonic behaviour of this ratio is preventing from distinguishing between fast and slow collapse scenarios\footnote{We note though, that also the ortho/para-H$_2$D$^+$ ratio will be affected by similar problems, as H$_2$D$^+$ will be converted to D$_2$H$^+$ over time.}.

In this work we revise this ratio by targeting six cores embedded in the well characterised L1688 and L1689 regions, two of the densest clouds of Ophiuchus, the first one considered to be more active than the second \citep{Nutter2006}. 
At a distance of 140 parsecs \citep{Ortiz2018}, and with a dense population of cores and young stellar objects \citep{Motte1998}, the Ophiuchus molecular cloud complex represents one of the closest active star-forming regions to the Solar system and then the ideal place to study the reliability of this ratio under different conditions.

The paper is organised as following: we first present the observational data and the analysis of the spectra (Section \ref{sect:observations}); in Section \ref{sect:simulations} we report the numerical simulations employed to interpret the observations; in Section \ref{sect:results} we compare observations and simulations and provide an estimate of the core ages. Finally, we present a comprehensive discussion and the conclusions (Section \ref{sect:discussion} and \ref{sect:conclusions}, respectively).

\section{Observational data}\label{sect:observations}
We have observed six cores within the Ophiuchus complex with the \textit{Atacama Pathfinder EXperiment} (APEX) \mbox{12-m} radio telescope \citep{Gusten2006} in both ortho-H$_2$D$^+$ ($J~=~1_{1,0} - 1_{1,1}$) at 372.42134 GHz and para-D$_2$H$^+$ ($J=1_{1,0} - 1_{0,1}$) at 691.66044~GHz \citep{Amano2005}. The sources have been chosen based on the properties of cores with already available observations of the ortho-H$_2$D$^+$ and para-D$_2$H$^+$ \citep{Parise2011,Vastel2004}, i.e. column density of molecular hydrogen N(H$_2$)$> 3\times 10^{22}$~cm$^{-2}$ and temperatures below $14$~K. A summary of the properties of the targeted cores are reported in Table~\ref{table:prop}.

Because of the very different line frequencies, the telescope resolution changes significantly, from $\sim 18''$ at $\sim 372.42$~GHz to $\sim9''$ at $691.66$~GHz. This would make the source size critical for correctly deriving the ratio of the two lines. In order to avoid this uncertainty, we have observed the cores with a single pointing centred on the maximum of the H$_2$ column density and the minimum of the dust temperature in the case of ortho-H$_2$D$^+$ ($1_{1,0} - 1_{1,1}$), and we covered the same area with a small on-the-fly map in the case of para-D$_2$H$^+$ ($1_{1,0} - 1_{0,1}$), so that the final extracted spectra are directly comparable.

The observations of ortho-H$_2$D$^+$  were performed on 27-28 April 2019 and on 6-8 June 2019 with the PI instruments LAsMA \citep{Gusten2008} and FLASH$^+$ \citep{Klein2014} under good weather conditions, with a precipitable water vapour (PWV) $\sim 0.5$~mm. The average system temperatures were in the range 539-899 K. We used a position-switch scheme, with carefully selected off-positions (\mbox{RA$_{2000}$=16h 32m 05.51s}, DEC$_{2000}$=-24d 52m 33.07s for cores 1 and 2, and RA$_{2000}$=16h 27m 46.355s, DEC$_{2000}$=-24d 31m 24.93s for the other sources), where the continuum emission has a local minimum in the \textit{Herschel} maps. The observations were organised in on-off loops of 20 seconds of integration, with regular chopper-wheel calibrations. Pointing and focus were performed as needed using Jupiter, NGC6073, IRAS15194 and IRC+10216. The total telescope time was typically 45 minutes per core, with an on-time of 20 minutes. The corrected antenna temperature was transformed to a main beam temperature ($T_\mathrm{mb}$) using a beam efficiency of $\eta_\mathrm{mb}=0.73$ and $\eta_\mathrm{mb} = 0.62$ for the LAsMA and FLASH$^+$ receivers, respectively. The final rms noise is in the range $30-50$~mK at a velocity resolution of $0.2$~km~s$^{-1}$. 
 
The ground-state line of para-D$_2$H$^+$ was observed with the SEPIA (band 9) receiver \citep{Hesper2005} on the following days: 27 April 2019, 23-24 May 2019, 02-03 June 2019, 07 July 2019, 28-29 July 2019, and 31 July 2019. The weather was exceptionally good, with PWV between $0.15$ and $0.3$~mm; the observations with sub-optimal PWV ($\gtrsim 0.5$~mm) were discarded, due to the high noise levels. The average system temperatures were in the range 1047-2164 K.
As mentioned, we observed small on-the-fly maps (\SI{18}{\arcsecond}$\times$\SI{18}{\arcsecond}) in order to cover the beam area of the telescope at $372$~GHz. The maps were observed following a zigzag pattern, with a dump time of one second; calibration was repeated after every complete execution. Pointing and focus were regularly performed on Jupiter, W-Hya, NGC6334I, Rx-Boo, and IRC+10216. We have applied a beam efficiency of $\eta_\mathrm{mb}= 0.52$ to convert the corrected antenna temperature to main beam temperature.
The final spectra for para-D$_2$H$^+$ have an rms noise in the range $20-70$~mK at a velocity resolution of $0.3$~km~s$^{-1}$.

\begin{figure*}
\centering
\includegraphics[scale=0.4]{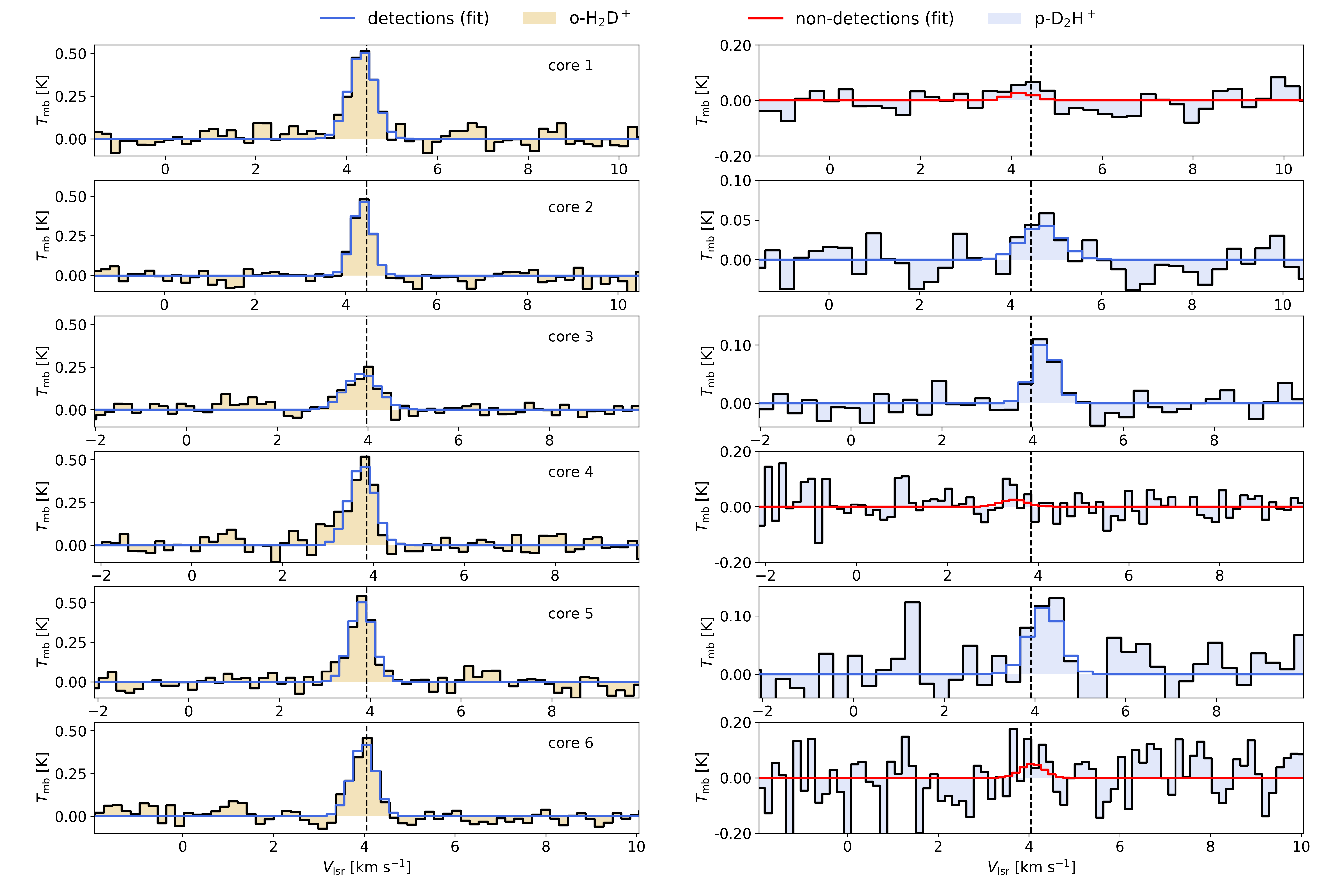}
\caption{Observed and modelled spectra of ortho-H$_2$D$^+$ (left) and para-D$_2$H$^+$ (right) for the six cores. Note that the temperature scale of the spectra is not the same in each panel. We have centred the spectra at the systemic velocity of ortho-H$_2$D$^+$ (vertical line) of each source. }\label{figure:spectra}
\end{figure*}

\subsection{Data analysis and fitting procedure} \label{sect:analysis}
The data were reduced using the \textsc{class} program of the \textsc{gildas} software\footnote{GILDAS is a software package developed by the Institute de Radio Astronomie Millim\'etrique (IRAM); see https://www.iram.fr/IRAMFR/GILDAS/}. After removing the spectra with clear problems, a baseline of degree-1 was subtracted around the line, and the spectra were finally averaged, weighting them on the noise level. For the SEPIA (band 9) observations, all positions were averaged as well, to extract the spectrum over the same area as for ortho-H$_2$D$^+$. 

We detected the ground-state line of ortho-H$_2$D$^+$  in the entire sample (six prestellar cores) with a signal-to-noise ratio (S/N) above 5$\sigma$. We also observed the ground state transition of the para-D$_2$H$^+$, obtaining two weak detections (S/N $\sim$ 3$\sigma$) and securing one detection (S/N $\sim$ 5$\sigma$).
For the other three sources we can only provide upper limits.   The spectra for both tracers are reported in Fig. \ref{figure:spectra}.

\begin{table*}
\centering
\caption{We report, the logarithm of the o-H$_2$D$^+$ column density, the full-width at half-maximum for the o-H$_2$D$^+$ spectra ($\Delta \upsilon_\mathrm{obs}$), the virial parameter, the Mach number $\mathcal{M}$, the p-D$_2$H$^+$ column density, the N(o-H$_2$D$^+$)/N(p-D$_2$H$^+$) ratio (OPR), and finally the fractional abundances with respect to H$_2$, $\mathrm{x}$(o-H$_2$D$^+$) and $\mathrm{x}$(p-D$_2$H$^+$), respectively. The full-width at half-maximum of p-D$_2$H$^+$ is not shown but is consistent, within the error, with the o-H$_2$D$^+$ values.  The quantities are calculated from the equations reported in the Appendix A by assuming $T_\mathrm{ex} = T_\mathrm{dust}$.}\label{table:results}
\setlength{\tabcolsep}{2pt}
\begin{tabular*}{\textwidth}{@{\extracolsep{\fill} }ccccccccc}
\hline
\hline

\#  &  log$_{10}$[N(o-H$_2$D$^+$)] & $\Delta \upsilon_\mathrm{obs}$ & $\alpha$  & $\mathcal{M}$ & log$_{10}$ [N(p-D$_2$H$^+$)] & OPR & $\log_{10}[\mathrm{x}$(o-H$_2$D$^+$)] & $\log_{10}[\mathrm{x}$(p-D$_2$H$^+$)]\\
core & [cm$^{-2}$]& [km s$^{-1}$] & & & [cm$^{-2}$] &&  & \\
\hline\\
1 &	 12.99$_{-0.08}^{+0.08}$ & 0.67$_{-0.11}^{+0.13}$ & 2.41$_{-0.46}^{+0.40}$ & 0.94$_{-0.19}^{+0.18}$ & $< 12.87$ & $> 1.32$ & -9.49$_{-0.13}^{+0.13}$ & $<$ -9.61\\
2 &    12.84$_{-0.09}^{+0.07}$ & 0.49$_{-0.08}^{+0.09}$ & 0.57$_{-0.10}^{+0.10}$ &  0.35$_{-0.17}^{+0.15}$ &12.50$_{-0.37}^{+0.29}$ & 2.19$_{-0.88}^{+0.63}$ & -9.89$_{-0.13}^{+0.13}$ & -10.23$_{-0.35}^{+0.35}$\\
3  &   12.80$_{-0.11}^{+0.09}$ & 0.84$_{-0.18}^{+0.21}$ & 0.60$_{-0.16}^{+0.12}$&   1.61$_{-0.24}^{+0.23}$ &12.90$_{-0.17}^{+0.17}$ & 0.81$_{-0.20}^{+0.14}$ & -10.17$_{-0.14}^{+0.14}$ & -10.07$_{-0.20}^{+0.20}$\\
4  &  12.96$_{-0.09}^{+0.10}$ & 0.66$_{-0.15}^{+0.16}$ & 1.17$_{-0.30}^{+0.25}$ &  0.94$_{-0.22}^{+0.29}$ &$< 12.74$ & $> 1.66$ & -9.86$_{-0.13}^{+0.13}$ & $<$ -10.08\\
5 &	 12.89$_{-0.08}^{+0.08}$ & 0.59$_{-0.11}^{+0.14}$ & 1.20$_{-0.29}^{+0.22}$ &  0.62$_{-0.19}^{+0.25}$ &12.70$_{-0.41}^{+0.31}$ & 1.40$_{-0.58}^{+0.42}$ & -9.94$_{-0.13}^{+0.13}$& -10.13$_{-0.37}^{+0.37}$\\
6 &   12.95$_{-0.08}^{+0.09}$ & 0.60$_{-0.12}^{+0.12}$ & 0.74$_{-0.16}^{+0.13}$ & 0.88$_{-0.21}^{+0.22}$ & $< 13.14$ & $> 0.65$ & -9.73$_{-0.13}^{+0.13}$ & $<$ -9.54\\
\hline
\hline
\end{tabular*}
\end{table*} 


Once the spectra of each core have been produced, to estimate the column densities of ortho-H$_2$D$^+$ and para-D$_2$H$^+$ we employed \textsc{MCWeeds} \citep{Giannetti2017}. \textsc{MCWeeds} allows to fit the observed spectrum with a number of algorithms (from simple maximum a posteriori estimates to Markov chain Monte Carlo), in order to obtain a robust estimate of the uncertainties. At every iteration, a synthetic spectrum is generated, under the assumption of local thermodynamic equilibrium (LTE, well justified by the high-density of the cores\footnote{The critical densities are $\sim 10^5$ cm$^{-3}$ and $\sim 5.6\times 10^5$ cm$^{-3}$, for ortho-H$_2$D$^+$ and para-D$_2$H$^+$, respectively.}), taking into account line opacity effects, and is then compared to the observed spectrum. Given that we have only two lines, performing a non-LTE analysis would not improve the quality of our inference because of the additional free parameters that would be needed for this model.
 
We have selected Markov chain Monte Carlo (MCMC) as the fitting method, to correctly evaluate the uncertainties in column densities, and propagate them numerically onto the ortho-H$_2$D$^+$/para-D$_2$H$^+$ ratio, and other derived quantities.
We have used 100000 total iterations, with a burn-in period of 10000 iterations, a delay for the adaptive sampling of 5000 iterations, and a thinning factor of 20. The traces were inspected to ensure convergence and independence of the samples.
The parameters that could be varied are six: column density of the species, excitation temperature, source size, line full-width at half-maximum, $v_{LSR}$, and a calibration factor. In this case, we have fixed the source size, assuming that it is extended compared to the beam; this means that the column densities that we obtain are beam-averaged. An extended emission region is supported by the results reported in previous works \citep{Vastel2006,Pagani2009_2,Parise2011}.
As we have only two transitions, we cannot independently constrain the excitation temperature $T_\mathrm{ex}$.  We have then approximated the excitation temperature with the dust temperature ($T_\mathrm{ex} = T_\mathrm{dust}$, see Table~\ref{table:prop}), assuming that gas and dust are coupled at the high central densities of the selected sources. While the dust temperatures can be influenced by an incomplete removal of the emission from the warmer, outer layers of the cores, the values that we obtain are low, reasonably compatible with cold, prestellar cores \citep{Caselli2008}. The effect of background and foreground is thus relatively small. 
We have inspected NH$_3$ data reported in literature on the same regions \citep{Friesen2009,Chitsazzadeh2014,Friesen2017}, and they are consistent, within the errors, with our $T_\mathrm{dust}$, which confirms that our assumption is reasonable. 
In Section \ref{sec:tex} we provide a comparison of the results obtained at different $T_\mathrm{ex}$ to assess the error introduced by our assumption.

Our final free parameters are then the column density, the line full-width at half-maximum, and $v_{LSR}$.
The priors we used are as follows: a normal distribution for the logarithm of the column density and $v_{LSR}$ with a $\sigma$ of 1.5 orders of magnitude, and $2$~km~s$^{-1}$, respectively. For the linewidth we used a truncated normal distribution, with $\sigma=1$~km~s$^{-1}$, truncated at $0.3$~km~s$^{-1}$ and $5$~km~s$^{-1}$.
Tests with different priors show that their exact form and parameters do not have a strong influence on the fitting results.  
 
Having a robust estimate of the uncertainty on the derived parameters is crucial for low signal-to-noise ratio observations. Since we are combining observations carried out in different moments and with different receivers, it is particularly important to account for calibration uncertainties, considering also that the column density is proportional to the integrated intensity of the line. For this we used a random factor, normally distributed, with $\sigma=5\%$, truncated at $\pm30\%$. The latter is turned into a multiplicative factor that we use to adjust the line intensity at each iteration, thus including it in the error budget. 


In the case of a non-detection, the upper limits of the column densities are derived by taking the value corresponding to the upper limit of the 95\% of the high probability density interval. The linewidth was fixed at the ortho-H$_2$D$^+$ value. 

We report in Fig.\ref{figure:spectra} the modelled spectra for the cores for both ortho-H$_2$D$^+$ and para-D$_2$H$^+$ (blue and red lines for detections and non detections, respectively), while in Table \ref{table:results} we summarise the results of our analysis together with the constrained physical quantities derived for the six cores (see Appendix A) and the relative errors obtained from our Bayesian analysis. The abundances are in line with previous studies \citep{Parise2011}, and the ortho-H$_2$D$^+$/para-D$_2$H$^+$ ratio falls in the range from 0.81 to 2.18.
The six targeted cores show observed temperatures and densities typical of prestellar cores ($T_\mathrm{dust}\sim 10-13$ K and $10^5 \leq n_\mathrm{core} \leq 10^6$ cm$^{-3}$, Table \ref{table:prop}), and the analysis of the spectral features suggests a dynamical scenario with the cores being gravitationally unstable ($\alpha\leq 2$ and $M > M_J$, see Appendix \ref{sect:appendix} for details on how these quantities have been calculated). This indicates that the energy support is not sufficient to prevent collapse and that the cores are likely to fragment. With the Mach number 
being close to one, we note that the effective Jeans mass could increase 
at most by a factor of $\sqrt{4/3}$ in the presence of turbulence \citep{klessen2004}. Even considering the effect of magnetic fields, it would require a magnetic field of at least  600-700 $\mu$G for the cores to be in virial equilibrium (for details on how this is calculated see \citealp{Redaelli2021}). 

 Such values in principle are not too far from the observed values of magnetic fields \citep{Crutcher19} even though there is a considerable spread, and it might be that the system is approximately virialized on the core scale. However, as we will show below (Section \ref{sect:discussion}) from the comparison with the dynamical models, an approximate virialization on the core scale does by no means imply that the system itself is long-lived, and indeed virialization on the core scale should naturally be expected even in models of a rapid collapse.

\section{Simulations}\label{sect:simulations}
To explore the physics of the observed cores and provide quantitative information on their dynamical state and temporal evolution we performed \textit{a priori} three-dimensional magneto-hydrodynamic simulations of star-forming filaments with on-the-fly non-equilibrium chemistry. Gravitationally bound cores are in fact embedded in thermally supercritical filaments as reported by studies based on high resolution observations \citep{Andre2010,Molinari2010} and in line with the hierarchical paradigm of solar-type star formation as the result of fragmentation processes within dense filaments.  We follow a rigorous approach by exploring different physical and chemical initial conditions, particularly focusing on the parameters that mostly affect the collapse of the filament and the formation of overdense regions. 
Different configurations of the magnetic fields can affect the fragmentation mode and the final outcome of the star formation process \citep{Seifried2015}, and parameters like the value of the cosmic-ray ionisation rate and the initial ortho-to-para H$_2$ ratio may drastically change the chemical evolution \citep{Sipila2015,Bovino2019}. 

In the following sections we report the details of our numerical setup and chemical model and the post-processing of our simulations, before comparing with observations.

\begin{table*}
\caption{Details of the performed simulations. We report the filament size, dust temperature, the filament mass $M_\mathrm{filam}$, the Mach number $\mathcal{M}$, filament central density $n_0$, initial magnetic field strength $B_0$ and the initial chemical parameters: H$_2$ ortho-to-para ratio and average grain size, $\langle a \rangle$. We set the instability parameter $(M/L)_\mathrm{fil}/(M/L)_\mathrm{cr}$= 3 for all the runs, with $(M/L)_\mathrm{cr} = 2c_s^2/G$, $c_s$ the thermal sound speed and $G$ the gravitational constant. Note that in our isothermal simulations we assume $T_\mathrm{dust} = T_\mathrm{gas}$.}\label{table:sim}
\setlength{\tabcolsep}{3pt}
\centering
\begin{tabular*}{\textwidth}{@{\extracolsep{\fill} } lccccccccccc}
\hline 
\hline
\# & Size & $T_\mathrm{dust}$ & $M_\mathrm{filam}$ &   $\mathcal{M}$  &  $n_0$ & $B_0$  &  B-field & OPR(H$_2$) & $\langle a \rangle$\\
core & [pc] & [K] & [M$_{\odot}$]& &  [cm$^{-3}$] & [$\mu$G] &  & & [$\mu$m] \\
 \hline
F1\_0035 & 1.6 & 15 & 120  & 5 & 7.6 $\times$ 10$^{4}$ &  40 & $\parallel$  & 0.1 & 0.035 \\
F1\_0035\_OPR & 1.6 & 15 & 120  & 5 & 7.6 $\times$ 10$^{4}$ &  40 & $\parallel$  & 0.001  & 0.035\\
F1\_01 & 1.6 & 15 & 120  & 5 & 7.6 $\times$ 10$^{4}$ &  40 & $\parallel$  & 0.1 & 0.1\\
F2\_01 & 1.6 & 10 & 80  & 1 & 5.0 $\times$ 10$^{4}$ &  40 & $\parallel$ & 0.1  & 0.1\\
F3\_01 & 1.6 & 10 & 80  & 1 & 5.0 $\times$ 10$^{4}$ &  40 & $\bot$ & 0.1 & 0.1\\
\hline
\hline
\end{tabular*}
\end{table*}

\subsection{Numerical setup}
The simulations presented in this work have been performed with the publicly available hydrodynamic code \textsc{gizmo} \citep{Hopkins2015}, descendant of \textsc{gadget2} \citep{Springel2005}. The code evolves the magneto-hydrodynamics equations for the gas including a constrained-gradient divergence-cleaning method \citep{Hopkins2016a,Hopkins2016b}, together with the gas self-gravity. For the purpose of this study, we equipped the code with an on-the-fly non-equilibrium chemical network, implemented via the public chemistry library \textsc{krome} \citep{Grassi2014}, similarly to \citet{Bovino2019}. In particular, in our simulations, we assume an isothermal equation of state for the gas, with the temperature set to 10~K or 15~K. These temperatures are in line with kinetic temperatures obtained from NH$_3$ for the same regions \citep{Friesen2009,Friesen2017}. 

The initial conditions of our simulations consist of a collapsing filament, modelled as a cylinder with a typical observed \citep{Arzoumanian2011} Plummer  like density profile $\mathbf{n(R) = n_0/[1+(R/R_\mathrm{flat})^2]^{p/2}}$, where $R$ is the cylindrical radius, $n_0$ is the ridge volume density, that is constant along the filament axis, $R_\mathrm{flat}$ is the characteristic radius of the flat inner part of the density profile, and $p$ the characteristic exponent. Following observed estimates \citep{Arzoumanian2011} we set $p = 2$ and $R_\mathrm{flat} = 0.033$~pc, which gives a mean filament  width of 3$\times R_\mathrm{flat} \sim 0.1$ pc. The setup follows previous works \citep{Seifried2015,Kortgen2018}. To avoid any spurious effect at the edges of the filament along its axis, we embed the cylinder in an exponentially decaying background, with the decay scale length set to the filament length $L_{\rm fil}=1.6$~pc. The box is large $L_\mathrm{box} = 2.4$ pc. 
We initialise the filament in a turbulent state, assuming a Burgers-like power spectrum that grows as $\propto k^{10}$ up to $\lambda_{\rm peak}=L_{\rm box}/6$ and then decays as $\propto k^{-2}$ \citep[see also][]{Kortgen2018}. Finally, we assume the box is permeated by a constant magnetic field $B_0=40\rm\, \mu G$ \citep{Seifried2015,Kortgen2018}.

In order to explore the parameter space of physical/chemical conditions of the observed prestellar cores and filaments, we perform a total of five runs, in which we vary different parameters. In detail, we consider a highly turbulent filament (F1), with Mach number 5 and ridge density 7.6$\times$10$^4$ cm$^{-3}$, in which we vary the average dust grain size $\langle a\rangle$ (0.0035~$\mu$m or 0.1~$\mu$m)\footnote{Processes like adsorption/freeze-out strongly depend on the grain-size ($\propto \langle a \rangle ^{-1}$). For this reason we varied this quantity by taking the standard 0.1~$\mu$m average size and the one obtained from a proper distribution \citep{Mathis1977} with a power-law index of 3.5 and $a_\mathrm{min}=10^{-7}$ and $a_\mathrm{max} = 2.5\times 10^{-5}$ cm, which provides an average size of $\langle a \rangle = 0.035$~$\mu$m.} and the initial H$_2$ ortho-to-para ratio (0.1 or 0.001), and two more gravitationally bound cases (F2 and F3) that start from a lower density ($n_0 = 5\times10^4$ cm$^{-3}$) and a lower temperature ($T_\mathrm{dust} = 10$ K), in which we assume the fiducial chemical parameters, i.e. $\langle a\rangle = 0.1$ and an H$_2$ ortho-to-para ratio of 0.1, and vary the magnetic field direction, i.e. parallel or perpendicular to the filament axis, respectively.
The choice of the initial H$_2$ ortho-to-para ratio is dictated by the few observations of this quantity: the conservative value of 0.1, lies in between typical values in diffuse clouds \citep{Neufeld1998,Crabtree2011} and indirect estimates obtained in dense clouds \citep{Trompscot2009}. We also report a case with 0.001, as recently estimated for star-forming regions \citep{Brunken2014}.

The filaments are ``collapsing filaments" with mass per unit length ($M/L$) three times the critical one \citep{Ostriker1964,Andre2010} and total mass of 80-120~M$_\odot$.
 The mass and spatial resolution of our simulations  are $10^{-4}$ M$_\odot$ and $\sim$20 AU, respectively, optimal to resolve the core structure.
  The parameters of the realised simulations are reported in Table~\ref{table:sim}. Possible caveats about the choice of our initial conditions are discussed in Section \ref{sect:caveats}.


\begin{figure*}
\begin{center}
\includegraphics[width=0.5\columnwidth,trim=0.3cm 0.3cm 0.3cm 0cm, clip]{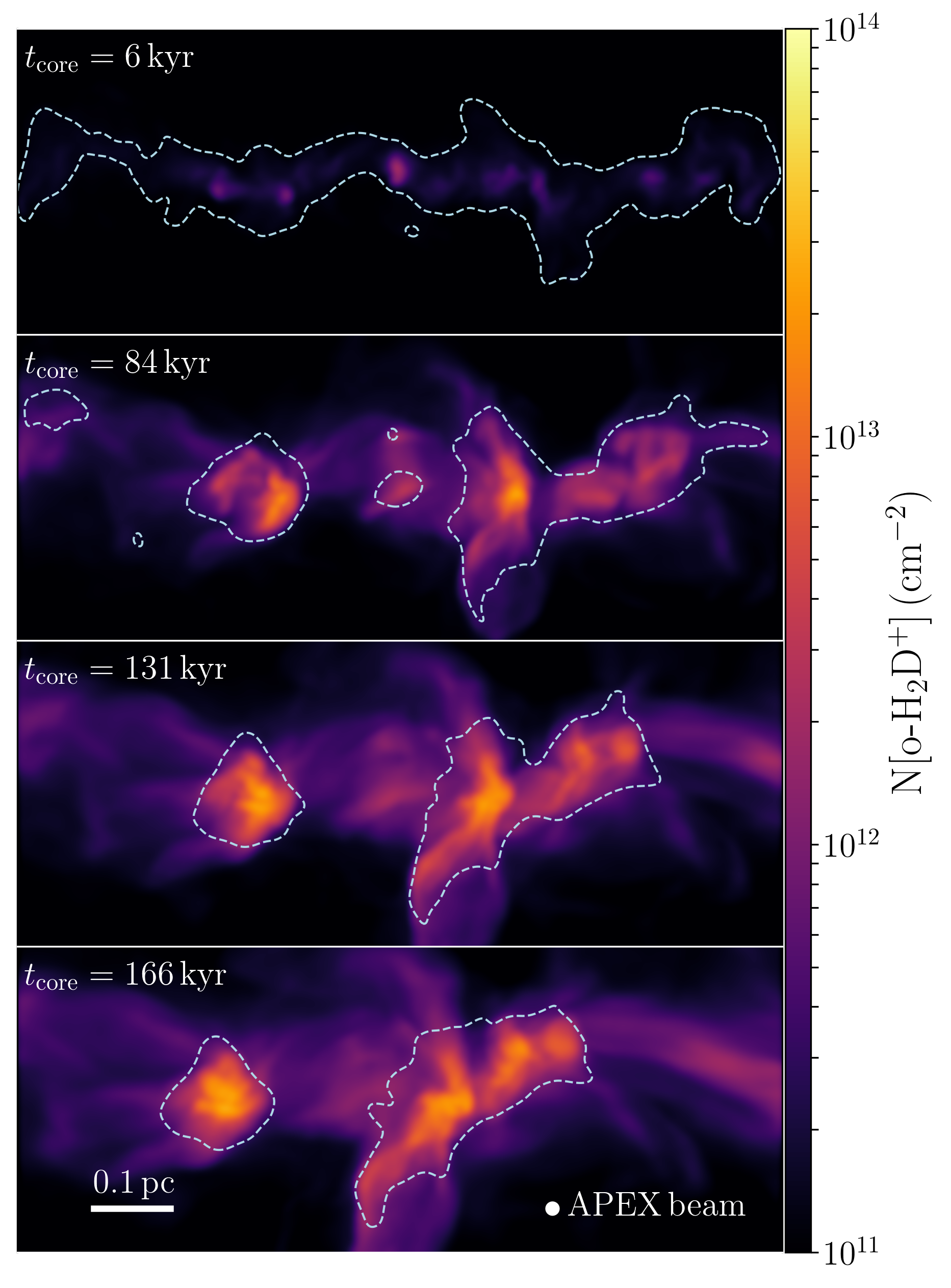}
\includegraphics[width=0.5\columnwidth,trim=0.3cm 0.3cm 0.3cm 0cm, clip]{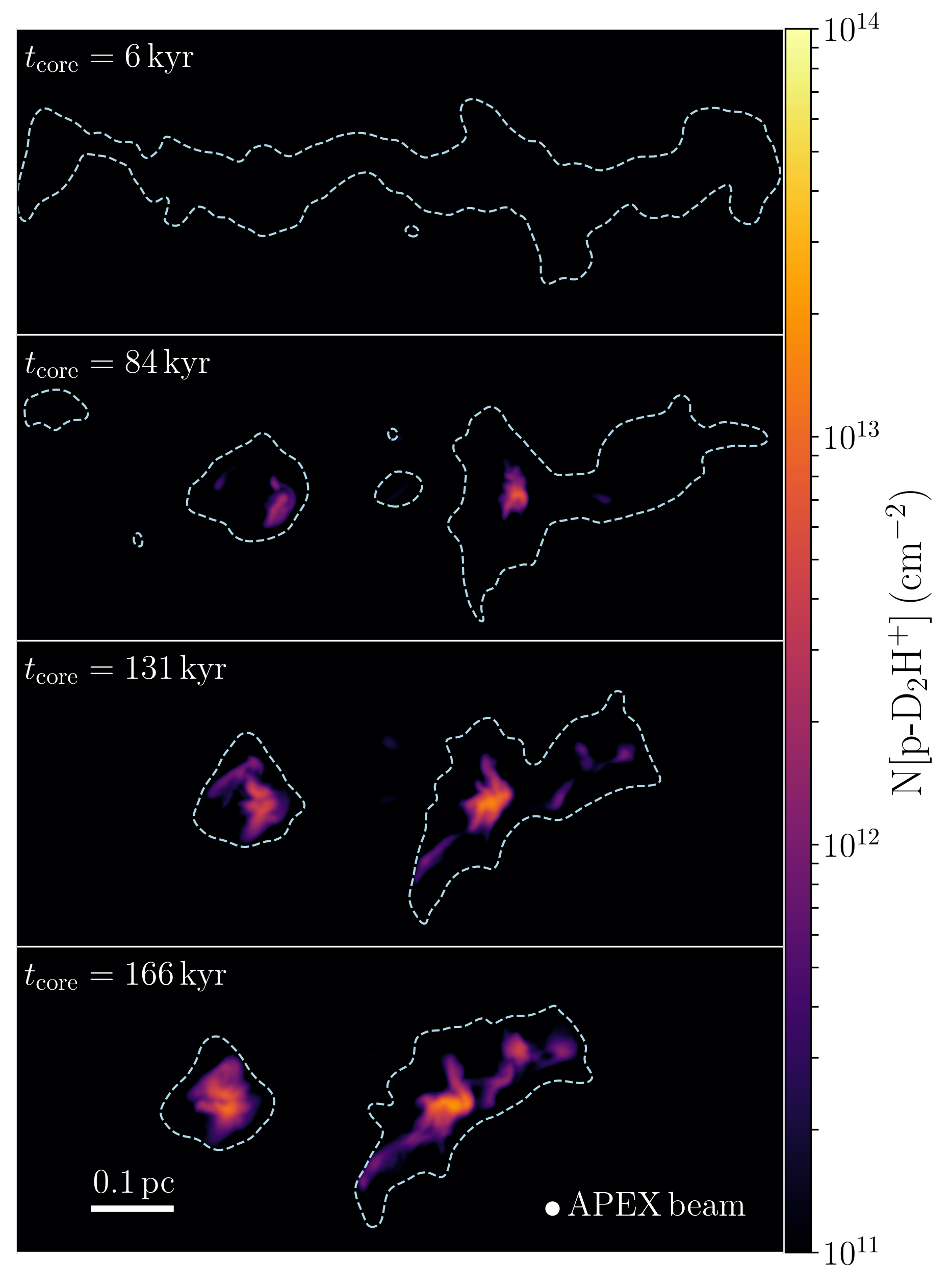}
\includegraphics[width=0.5\columnwidth,trim=0.3cm 0.3cm 0.3cm 0cm, clip]{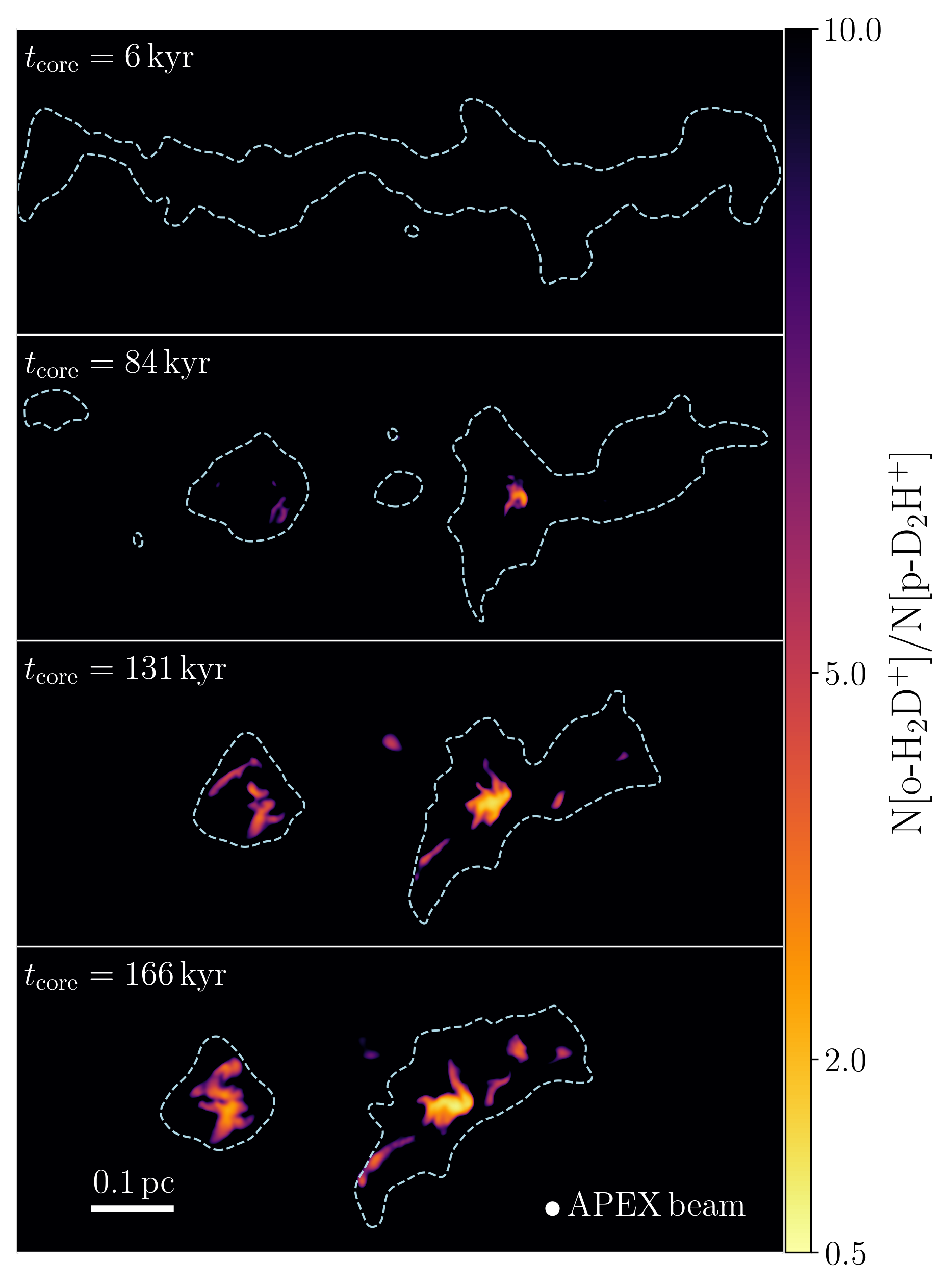}
\includegraphics[width=0.5\columnwidth,trim=0.3cm 0.3cm 0.3cm 0cm, clip]{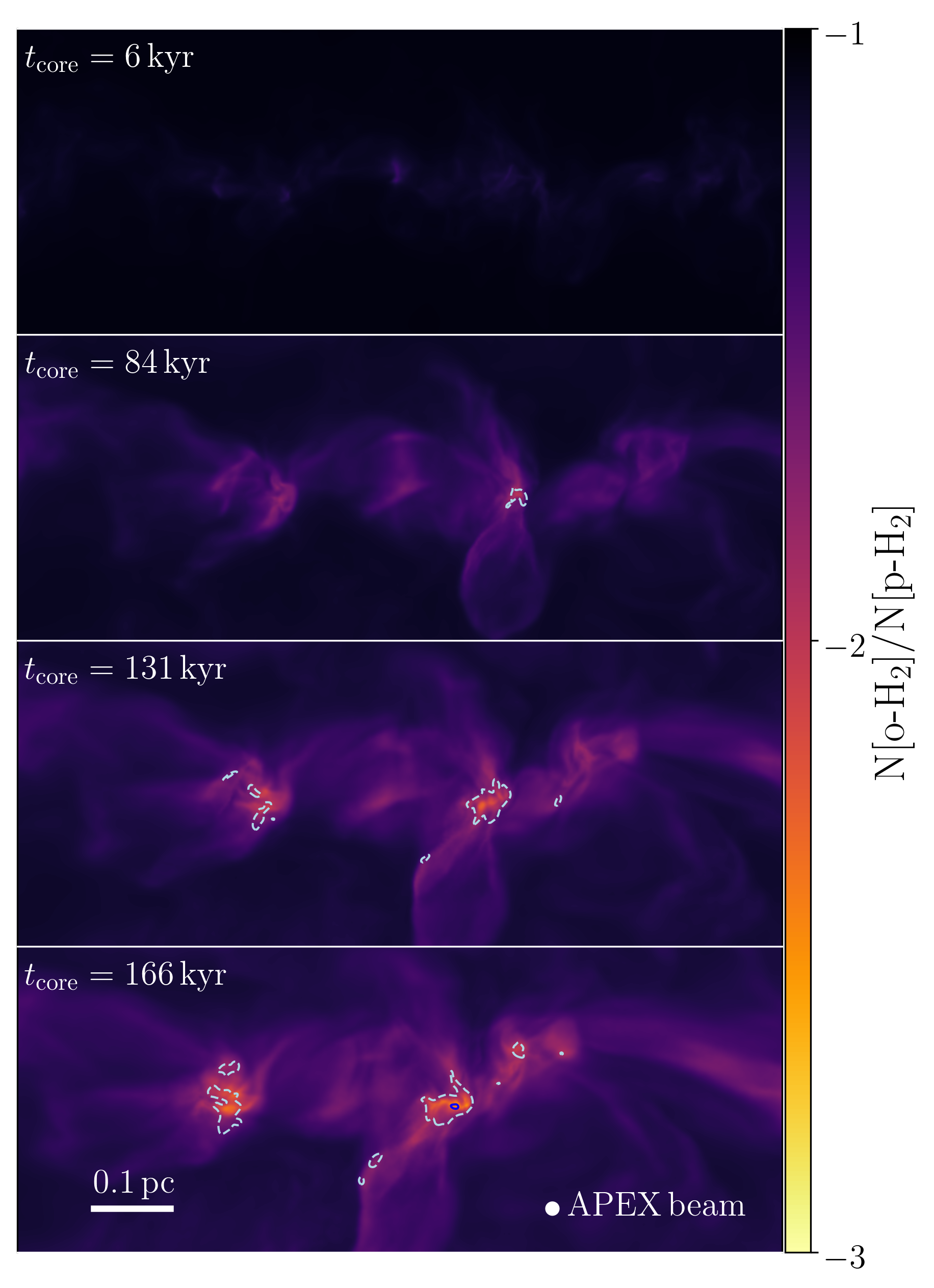}
\end{center}
\caption{From left: map of N(o-H$_2$D$^+$), N(p-D$_2$H$^+$), the N(o-H$_2$D$^+$)/N(p-D$_2$H$^+$) ratio, and the o/p H$_2$ from the reference run F1\_01. Each  panel reports the result for different snapshots in time. Contour regions represent H$_2$ total column density above 10$^{22}$ cm$^{-2}$ (cyan) in the first three panels, while the last panel shows overlaid the  N(o-H$_2$D$^+$)/N(p-D$_2$H$^+$), where the dashed contour has a value of 5 and solid contour  1. As a reference, we also report the APEX beam used in the observations (\SI{18}{\arcsecond}), the spatial scale in parsec, and the age of the core ($t_\mathrm{core}$). The maps have been convolved with the APEX beam of FWHM \SI{18}{\arcsecond} and \SI{9}{\arcsecond} for the ortho-H$_2$D$^+$ and the para-D$_2$H$^+$, respectively. The total H$_2$ column density distribution is shown after convolution with a typical Herschel beam of \SI{36}{\arcsecond}. The ratio is calculated on a pixel-to-pixel basis.}\label{fig:simmaps}
\end{figure*}

\subsection{Chemical model}
The chemical network employed in this study includes 134 species, with a total of 4616 reactions, that are self-consistently evolved for each gas particle/cell in the simulation at every step of the run \citep{Bovino2019}. We include N-C-O  bearing species up to three atoms (e.g. N$_2$H$^+$, HCO$^+$ among many others), with spin isomers of the main species (ortho-, para-, and meta-) and isotopologues (all the deuterated species, e.g. N$_2$D$^+$, DCO$^+$ etc.). 
We have built this network performing benchmarks towards published works \citep{Sipila2015,Kong2015} and including latest updates \citep{Hugo2009,Pagani2009_2}\footnote{We note that in \citet{Sipila2017} some of the rates involving the H$_2$ + H$_3^+$ system have been updated by considering not only the ground state of the molecules but also the rotational excited states. However, the changes can be considered negligible as they are within a few percent at the low-temperatures ($T<15$ K) of interest in this work.}. The freeze-out of heavy species like CO and N$_2$ is followed time-dependently \citep{Bovino2019}, together with the thermal and non-thermal desorption. We consider dust grains to be silicates with a typical specific density, $\rho_0 = 3$ g cm$^{-3}$, and a dust-to-gas mass ratio of $\mathcal{D} =  7.09 \times 10^{-3}$ as reported in previous works \citep{Kong2015}.
The chemical species are initialised by following \citet{Bovino2019}, which considers a fully molecular stage, but we note that changing the initial chemical abundances has a very minor effect on the evolution \citep[see for instance][]{Kong2015}. The cosmic-ray ionisation flux per hydrogen molecule  has been fixed to $\zeta_2 = 2.5 \times 10^{-17}$ s$^{-1}$, typical values in the dense regions of molecular clouds \citep[e.g.][]{vanderTak2000,Padovani2018,Bovino2020}. Larger values have been also reported \citep{Ivlev2019}, and would in general favour the formation of deuterated species. It is worth noting that this is one of the most uncertain parameters and its effect on deuteration has been already reported in several previous works \citep[e.g.][]{Kong2015,Kortgen2018,Bovino2019}. In \citet{Kortgen2018}, the authors needed a factor of 5 difference in the cosmic-ray ionization rate to see a relevant difference in the deuteration. \citet{Bovino2019}, reported a difference in the timescale of about 1.4 to reach the same deuteration fraction when going from 2.5$\times 10^{-17}$ to 1.3$\times 10^{-17}$ s$^{-1}$, while this becomes more relevant (up to a factor of 4 in the timescale) when employing a very high cosmic-ray ionization rate (e.g. 2.5$\times 10^{-16}$ s$^{-1}$). Our choice is very conservative and guarantees a certain robustness in the final timescale estimates, which in the worst case can be considered as upper limits, due to the low value we employ.

\subsection{Overall evolution}
We evolve the filaments up to roughly 300 kyr and focus on the ortho-H$_2$D$^+$/para-D$_2$H$^+$ (ratio between column densities). 
An example of our simulation results is reported in Fig. \ref{fig:simmaps}, which show the reference case F1\_01: a massive (120 M$_\odot$), highly turbulent filament ($\mathcal{M}=5$), with $\langle a\rangle$ = 0.1 $\mu$m, $T = 15$~K, and initial H$_2$ OPR of 0.1.  
The left panels show the time evolution of the ortho-H$_2$D$^+$ and para-D$_2$H$^+$ column densities after convolution with the APEX beam point spread functions of FWHM \SI{18}{\arcsecond} and \SI{9}{\arcsecond}, respectively. 
In the following analysis, the evolution is shown as a function of the core time $t_\mathrm{core}$, which represents the time passed since the formation of the core at $t_\mathrm{form}$, the latter measured from the beginning of the simulation. The last important time-scale in the analysis is the age of the core $t_\mathrm{age}$, corresponding to the time at which we compare our simulations with the observations. The latter is taken at the minimum o-H$_2$D$^+$/p-D$_2$H$^+$ ratio reached in the simulations, which in most cases corresponds to the sink formation time.

To compute the column densities, we select gas particles within a cylinder of radius 0.1~pc, with the axis aligned to the filament, and integrate their species local density along the line-of-sight, over the corresponding kernel size. In this way, if observations are strongly affected by line-of-sight contamination, our results are consistently mimicking this effect.

As clear from the figure, once the cores form and during their early evolution, the ortho-H$_2$D$^+$ column density is very compact (e.g. $t_\mathrm{core}\sim 6$ kyr), and shows values around 10$^{12}$ cm$^{-2}$. The column density starts to increase quickly with time, covering the entire filament when reaching $\sim$131 kyr. We know from observations (e.g. \citealp{Pillai2012,Friesen2014}) that the ortho-H$_2$D$^+$ emission is very extended. This is resembled by our simulations, which indeed show quite extended ortho-H$_2$D$^+$. At later times, densities are high enough to start the conversion of ortho-H$_2$D$^+$ into other isotopologues. On the other hand, the para-D$_2$H$^+$ is much more compact, and need at least 81 kyr to reach  values around 10$^{11}$-10$^{12}$ cm$^{-2}$ in the very central region of the cores (less than 0.05~pc). This is due to the fact that to form \mbox{para-D$_2$H$^+$} a certain amount of H$_2$D$^+$ should be in place. Over time, the amount of para-D$_2$H$^+$ increases and covers the entire core (at a scale of 0.1~pc), resulting in ortho-to-para ratios around or below unity, as can be clearly observed in the middle-right panel, where we show  the ortho-H$_2$D$^+$/para-D$_2$H$^+$ ratio, overlaid the total column density (cyan contours). The specific ratio decreases over time (from top to bottom panel), and reaches typical observed values (0.7-1.0) in roughly 166 kyr since the formation of the cores, providing a clear evidence of the short lifetimes of the observed cores. Low ratios are observed in regions as large as 0.05-0.08~pc, triggered by the para-D$_2$H$^+$ evolution. 

\begin{figure*}
\begin{center}
\includegraphics[scale=0.5]{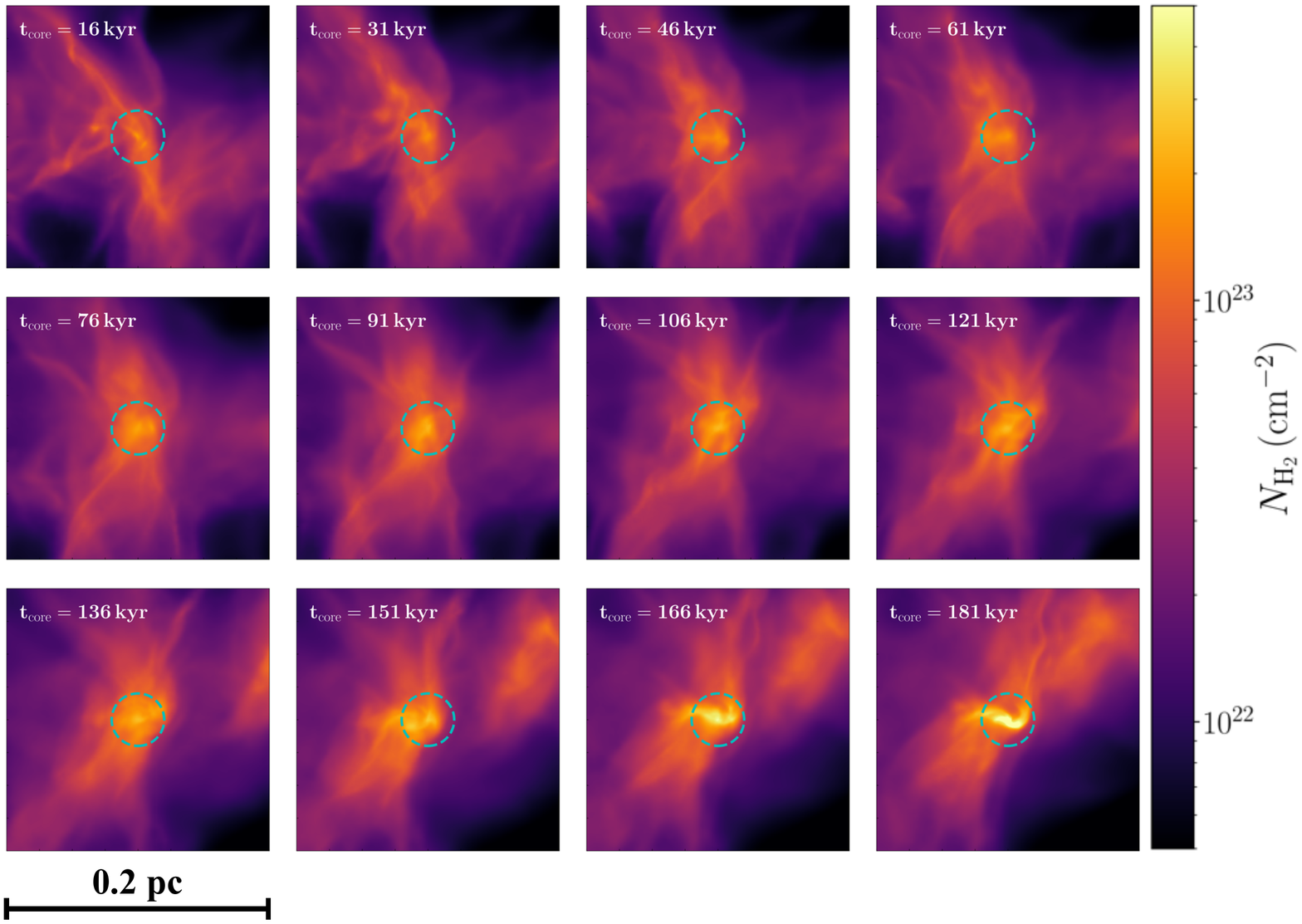}
\end{center}
\caption{Zoom on the core F1\_01\_1  reported in Fig. \ref{fig:simmaps} within the filament (core on the right, see also Table \ref{tab:coreprop}). The time evolution start from $t_\mathrm{core}$=16 kyr and goes up to the time at which a sink particle (i.e. a protostellar object) is formed, $t_\mathrm{core}$= 181 kyr. The dashed circle represents an area of 0.02 pc (\textit{Herschel}-beam) centered in the column density peak, which is the region that defines the core and over which we average to calculate all the core quantities reported in Table \ref{tab:coreprop}. Note that thes maps are not convolved.}\label{fig:simmaps2}
\end{figure*}

In the last panel of Fig. \ref{fig:simmaps} we report the ortho-/para-H$_2$ ratio with overlaid the ortho-H$_2$D$^+$/para-D$_2$H$^+$. The values of ortho-/para- H$_2$, as expected, decrease very quickly once the cores are formed. At 6 kyr the ratio is still close to the initial one (0.1) while decreases quickly when the gas becomes denser, reaching values below 10$^{-2}$ at about 166 kyr, with central regions peaking at 10$^{-3}$. In the same plot, we can appreciate how the ortho-H$_2$D$^+$/para-D$_2$H$^+$ (dashed contour value of 5, solid contour value of 1) probes very well, particularly on scales of 0.02 pc, the ortho-/para-H$_2$. This confirms that the ortho-H$_2$D$^+$/para-D$_2$H$^+$ ratio can be used as a proxy to determine the elusive ortho-/para-H$_2$ ratio. Further analysis on this specific point will be reported in a forthcoming paper.

Finally, a zoom into core F1\_01\_1 (the right one in Fig.~\ref{fig:simmaps}) is reported in Fig.~\ref{fig:simmaps2}, where we show its time evolution up to the formation of the sink particle ($t_\mathrm{core} = 181$ kyr). From this plot we can appreciate the highly dynamical evolution of the core. The central region of the core is collapsing very fast in the first 30 kyr of evolution, with accretion of material from filament scales. After this phase, it experiences a quasi-static evolution for at least 50-60 kyr. At about $t_\mathrm{core}=106$ kyr, the core starts to collapse with a continous flow of material from the surroundings, until reaching the formation of a protostellar object (sink particle) around 181 kyr, where also some rotation is well visible. This example shows that the evolution of the cores is highly non-linear and explain why the formation of the cores within certain simulations is taking hundreds of kyr.

The detailed analysis of the cores formed within the filaments will be discussed in the next Section.

\section{Comparison simulations - observations}\label{sect:results}
As we are mainly interested in the identification of the cores formed within the filaments and specifically in the time evolution of the ortho-H$_2$D$^+$/para-D$_2$H$^+$, we report in the following sections the detailed description of how this is performed and present a detailed comparison with the observational data.

\begin{table*}
\centering
\caption{Properties of the synthetic cores calculated at $t_\mathrm{age}$, i.e. at the minimum o-H$_2$D$^+$/p-D$_2$H$^+$. The properties are calculated by assuming an $R_\mathrm{eq}$ = 0.02 pc, which is an average of the effective size of the observed cores. We report the core mass $M_\mathrm{core}$, volume density $n_\mathrm{core}(t_\mathrm{age})$, ortho-to-para ratio, mach number ($\mathcal{M}$), and virial parameter ($\alpha$). We also report the abundances of o-H$_2$D$^+$ and p-D$_2$H$^+$, and the relevant timescales: the core formation time, $t_\mathrm{form}$, i.e. the time at which the core is formed, the core age $t_\mathrm{age}$, the free-fall time $t_\mathrm{ff}^0$ calculated at the $n_\mathrm{core}(t_\mathrm{form})$, the ratio between the age and the dynamical time $t_\mathrm{age}/t_\mathrm{ff}^0$, which gives an idea on the collapse speed, and the ambipolar diffusion time calculated by taking the electron fraction at the core formation time $t_\mathrm{form}$.}\label{tab:coreprop}
\setlength{\tabcolsep}{3pt}
\scriptsize
\begin{tabular*}{\textwidth}{@{\extracolsep{\fill} }lccccccccccccc}
\hline 
\hline
\# & $\log_{10}[n_\mathrm{core}(t_\mathrm{form})]$ & $M_\mathrm{core}$ &  $\log_{10}[n_\mathrm{core}(t_\mathrm{age})]$ & o-H$_2$D$^+$/p-D$_2$H$^+$ & $\mathcal{M}$ & $\alpha$ & $\mathrm{x}$(o-H$_2$D$^+$) & $\mathrm{x}$(p-D$_2$H$^+$) & $t_\mathrm{form}$ & $t_\mathrm{age}$ & $t_\mathrm{ff}^0$ & $t_\mathrm{age}/t_\mathrm{ff}^0$ & $t_\mathrm{AD}^0$\\
core  & [cm$^{-3}]$ & [M$_{\odot}$] & [cm$^{-3}$] & & & & $\log_{10}$ & $\log_{10}$ & [kyr]  &[kyr] & [kyr] & & [kyr] \\
 \hline
F1\_0035\_1      & 5.79 & 1.06 & 6.11 & 1.19 & 0.69  &  1.23 & -9.34 & -9.42  & 44  & 132 & 42 & 3.1 &640\\
F1\_0035\_2      & 5.79 & 1.15 & 6.14 & 0.94 & 0.71  &  1.17 & -9.42 & -9.40  & 44  & 215 & 42 & 5.0 &649\\
F1\_0035\_OPR\_1 & 5.79 & 1.33 & 6.21 & 0.79 & 0.91  &  1.31 & -9.94 & -9.84  & 44  & 166 & 42 & 3.9 &392\\
F1\_0035\_OPR\_2 & 5.79 & 0.97 & 6.07 & 1.57 & 0.71  &  1.38 & -9.82 & -10.02 & 44  & 176 & 42 & 4.1 &397\\
F1\_01\_1        & 5.79 & 1.33 & 6.21 & 0.94 & 0.91  &  1.32 & -9.75 & -9.72  & 44  & 166 & 42 & 3.9 &584\\
F1\_01\_2        & 5.79 & 1.00 & 6.08 & 3.18 & 0.70  &  1.33 & -9.66 & -10.17 & 44  & 185 & 42 & 4.3 &590\\
F2\_01\_1        & 5.74 & 1.06 & 6.11 & 2.22 & 1.26  &  1.92 & -9.43 & -9.77  & 213 & 102 & 45 & 2.2 &545\\
F2\_01\_2        & 5.70 & 0.62 & 5.88 & 2.63 & 0.42  &  1.38 & -9.36 & -9.78  & 249 & 75  & 47 & 1.6 &571\\
F2\_01\_3        & 5.70 & 0.84 & 6.00 & 2.66 & 0.91  &  1.39 & -9.46 & -9.89  & 196 & 108 & 47 & 2.3 &561\\
F2\_01\_4        & 5.70 & 0.57 & 5.84 & 3.74 & 0.31  &  0.98 & -9.38 & -9.96  & 256 & 69  & 47 & 1.5 &550\\
F2\_01\_5        & 5.70 & 0.73 & 5.95 & 2.40 & 0.37  &  0.81 & -9.37 & -9.76  & 223 & 102 & 47 & 2.1 &549\\
F3\_01\_1        & 5.74 & 0.66 & 5.90 & 1.84 & 0.51  &  1.05 & -9.44 & -9.70  & 294 & 36  & 45 & 0.8 &1057\\
F3\_01\_2        & 5.74 & 0.69 & 5.92 & 1.46 & 0.98  &  1.83 & -9.70 & -9.86  & 245 & 65  & 45 & 1.4 &453\\
F3\_01\_3        & 5.74 & 0.65 & 5.90 & 1.73 & 0.56  &  1.13 & -9.50 & -9.74  & 278 & 52  & 45 & 1.1 &801\\
F3\_01\_4        & 5.71 & 0.49 & 5.77 & 2.24 & 0.55  &  1.51 & -9.42 & -9.77  & 307 & 24  & 46 & 0.5 &1295\\
F3\_01\_5        & 5.74 & 0.63 & 5.88 & 1.76 & 0.89  &  1.78 & -9.50 & -9.75  & 282 & 43  & 45 & 0.9 &539\\
\hline
\hline
\end{tabular*}
\end{table*}

\subsection{Core identification in the simulated filaments}
In order to identify the prestellar cores in the simulated filaments, we first compute 2D column density maps of H$_2$, N(H$_2$), starting from the last snapshot in time  and moving back to the beginning\footnote{Being our simulations isothermal, the column density is directly proportional to the flux, and the identification procedure is therefore similar to the one followed for real observations.}. At each step, we estimate the average $\langle \mathrm{N}$(H$_2$)$\rangle$ in the filament map from a rectangular region of size $L_{\rm fil}\times0.2$ pc centred on the filament (in order to remove the low-density background from the analysis), and select all the pixels showing an overdensity \mbox{N(H$_2$)/$\langle$N(H$_2$)$\rangle > 10$}. Then, we apply a connected component analysis \citep{Fiorio1996} on the filtered pixels, labelling the identified regions, that represent our candidate cores. To avoid possible artefacts and unrealistic cores, we remove all regions with an axis ratio larger than 5 or smaller than 1/5, and those composed of less than 10 pixels. The remaining structures are then accepted as actual cores, if they are connected to their descendant via a minimum separation criterion, i.e. the cores are discarded if their centres are separated by more than $2\times \max\{r_\mathrm{prog},r_\mathrm{desc}\}$, where $r_\mathrm{prog}$ and $r_\mathrm{desc}$ are the progenitor and the descendant core effective sizes, respectively. If a core is the result of a merger, only the most massive progenitor is tracked back in time. Finally, for each core history, we estimate the average H$_2$ column density, the gas number and electron densities, and the effective size. In addition, we also compute column density maps of ortho-H$_2$D$^+$,  and para-D$_2$H$^+$, from which we measure the values convolved with the point spread function in the corresponding cores, in an observer-like fashion, by assuming a beam size of \SI{18}{\arcsecond} for the first one, and \SI{9}{\arcsecond} for the latter, and averaging over the ortho-H$_2$D$^+$ beam, according to the observations procedure. 
The H$_2$ column density has been convolved with a typical \textit{Herschel} beam of  \SI{36}{\arcsecond}, to match the angular resolution of the observations. 
All the quantities reported in this paper have been beam-averaged. 

\begin{figure*}
\includegraphics[scale=0.5]{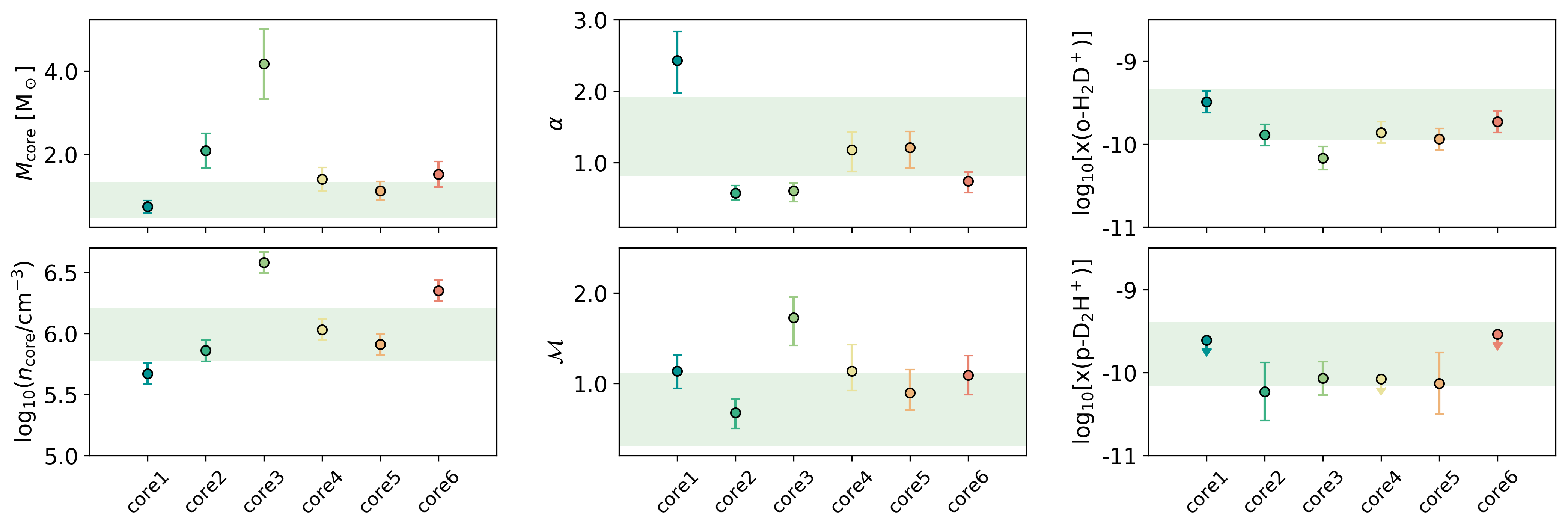}
\caption{Physical properties of the simulated and observed cores. We report the core mass and volume density (left columns), the virial parameter and Mach number (center), calculated from the formulae reported in Appendix A. In the right column we report also the fractional abundances of the two tracers in logarithmic scale, $\mathrm{x}$(o-D$_2$H$^+$), and $\mathrm{x}$(o-H$_2$D$^+$). Observed cores are reported as circles, while the green shaded area represents the range of values obtained for the simulated cores identified in our filaments (properties reported in  Table \ref{tab:coreprop}).}\label{fig:simcomp}
\end{figure*}

The analysis of the entire population of cores found in our simulations is reported in Table \ref{tab:coreprop}. The synthetic cores (sixteen in total), formed in the filaments, share similar physical properties with the observed ones, as shown in Fig. \ref{fig:simcomp}, with masses in between 0.5~--~1.3~M$_\odot$, densities around $5.8\times 10^{5}$ -- $1.6\times 10^{6}$ cm$^{-3}$ and transonic Mach numbers up to 1.2. The virial parameters estimated on the core scale lie in between 0.8 -- 1.9, indicating cores which are subvirial and/or virialised. The cores need an average time between 45-300 kyr to form ($t_\mathrm{form}$ in Table \ref{tab:coreprop}), depending on the configuration of the magnetic field and the turbulence level. In Fig. \ref{fig:simcomp} and Table \ref{tab:coreprop} we also report the fractional abundances of the two tracers. The comparison between the observed values and the synthetic ones shows a very good agreement, confirming that our simulations are accurate and provide a realistic representation of the observed regions.

By inspecting the dynamical state of the cores we can retrieve information on the evolution over time of physical quantities like the virial parameter. 
We note that all the cores in our simulations show evidence of fast collapse in time, which is not reflected in the estimated virial parameter at $t_\mathrm{age}$. 
This confirms that estimates of the virial parameter from observations may not actually allow to distinguish between rapid and slow collapse, as on the core scale subvirial or virialized cores would be expected in both scenarios; however, when considering larger scales including the envelope, one would expect a supervirialized configuration in case of rapid collapse, and it then becomes relevant on which scale the virial parameter is measured \citep[see also][]{Gomez2021}. 
We also estimate the average magnetic field strength along the line-of-sight for the synthetic cores. These are of the order of 70-300 $\mu$G, and fall in the range of values obtained from observational estimates of Ophiuchus \citep{Soam2018,Pattle2021}. Overall, within the approximations we have done in our simulations, and taking into account the errors coming from observations, we can state that our synthetic cores are well representing the observed ones by matching both physical and chemical constraints. However, we do not exclude that different dynamical situations can lead to similar results. In fact, a certain degree of degeneracy in the final comparison is unavoidable and large-scale simulations, which are not currently available, are the only viable tool to reduce the errors introduced by the choice of the initial conditions. In Section \ref{sect:caveats} we discuss different scenarios and what is the effect of a slow-collapse on our results.

\subsection{Cores time evolution and timescales}
\begin{figure*}
\centering
\includegraphics[width=0.9\textwidth]{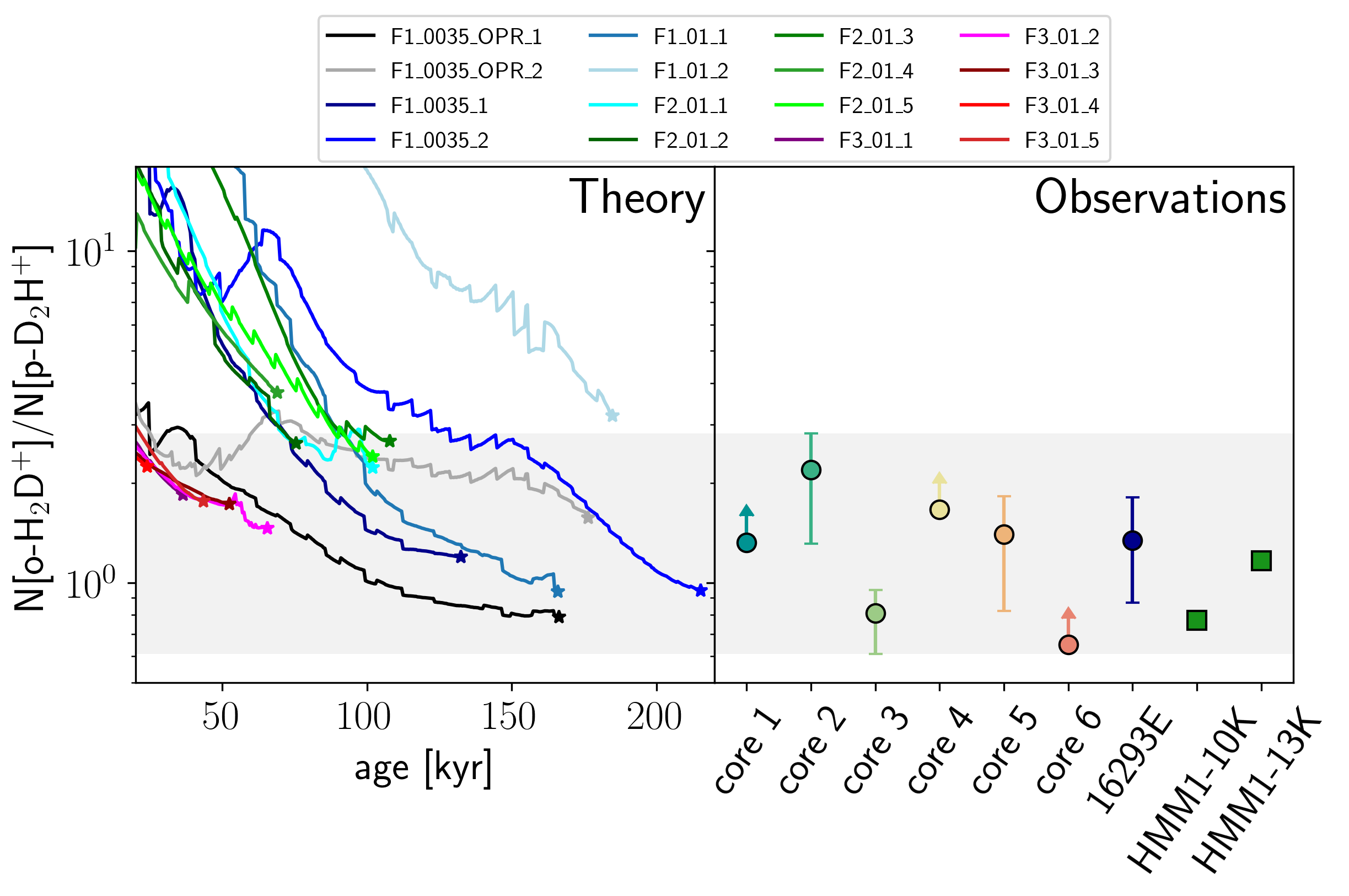}
\caption{Comparison of the N[o-H$_2$D$^+$]/N[p-D$_2$H$^+$] ratio between observations and simulations. The different curves in the left panel refer to the time evolution of the cores identified within the simulations with a different set of physical and chemical conditions, see Table \ref{table:sim} and  Table \ref{tab:coreprop} for details. Note that some cores are the result of merging and dissolving processes, which produce the visible oscillations in the curves. The stars represent $t_\mathrm{age}$, i.e. the time at which the minimum o-H$_2$D$^+$/p-D$_2$H$^+$ ratio is reached for each core and the point where we perform the calculations of the properties reported in Table \ref{tab:coreprop} and the comparison with the observed cores.
The right panel shows the observed values: the ones labeled 1-6 represent our APEX data, 16293E (blue circle) refers to previous observations of the two studied isotopologues \citep{Vastel2004} and HMM1-10K and HMM1-13K (green squares) are taken from observations in the L1688 cloud \citep{Parise2011}. The grey shaded area represents the range of observed values with secure detection, considering their uncertainties.}\label{fig:ratio}
\end{figure*}

In Fig. \ref{fig:ratio}, we show the time-evolution of the ortho-H$_2$D$^+$-to-para-D$_2$H$^+$ ratio of each core up to the minimum reached, by setting the zero-time at the first appearance of the core ($t_\mathrm{form}$). A downward trend is clearly visible, confirming that this ratio evolves with time and it is sensitive to changes in density during the collapse, which implies the conversion of H$_2$D$^+$ into D$_2$H$^+$  
 \citep{Flower2004,Pagani2013}. 
Even if not reported in the figure, some of the cores show an increasing trend after the minimum, likely due to the expected conversion of D$_2$H$^+$ into D$_3^+$ with time \citep{Flower2004} and most of them end up in a protostar, computationally defined by a sink particle.
In the same plot we also report the observed values (right panel and grey shaded area), including the only existing data from previous observations, i.e. the 16293E \citep{Vastel2004} and the HMM1 \citep{Parise2011} prestellar cores.  It is clear from the results that the observed values of the ortho-to-para ratio can be reached in a time no longer than 200 kyr after the formation of the cores, for the set of initial conditions employed here.  For the lower limits provided in the same figure, the time represents an upper limit to the age of the cores. The non-detection of para-D$_2$H$^+$ for three of the observed cores suggests that those sources are at a very early stage of the collapse, and H$_2$D$^+$ did not have time to be converted into D$_2$H$^+$ \citep{Flower2004}. Core 1, represents a scenario of relatively slow-collapse, being this core the one with the lowest density and the largest virial parameter (see Tables \ref{table:results} and \ref{table:prop}). However, without additional constraints (observational and theoretical) it is very difficult to  firmly distinguish between a slow collapse, an early stage of collapse, or a quasi-static temporary stage, and as we discussed the virial parameter alone is not a good indication of the dynamical state of the cores.

\subsection{Implications of observational uncertainties for the comparison with simulation}\label{sec:tex}
\begin{figure}
\centering
\includegraphics[scale=0.45]{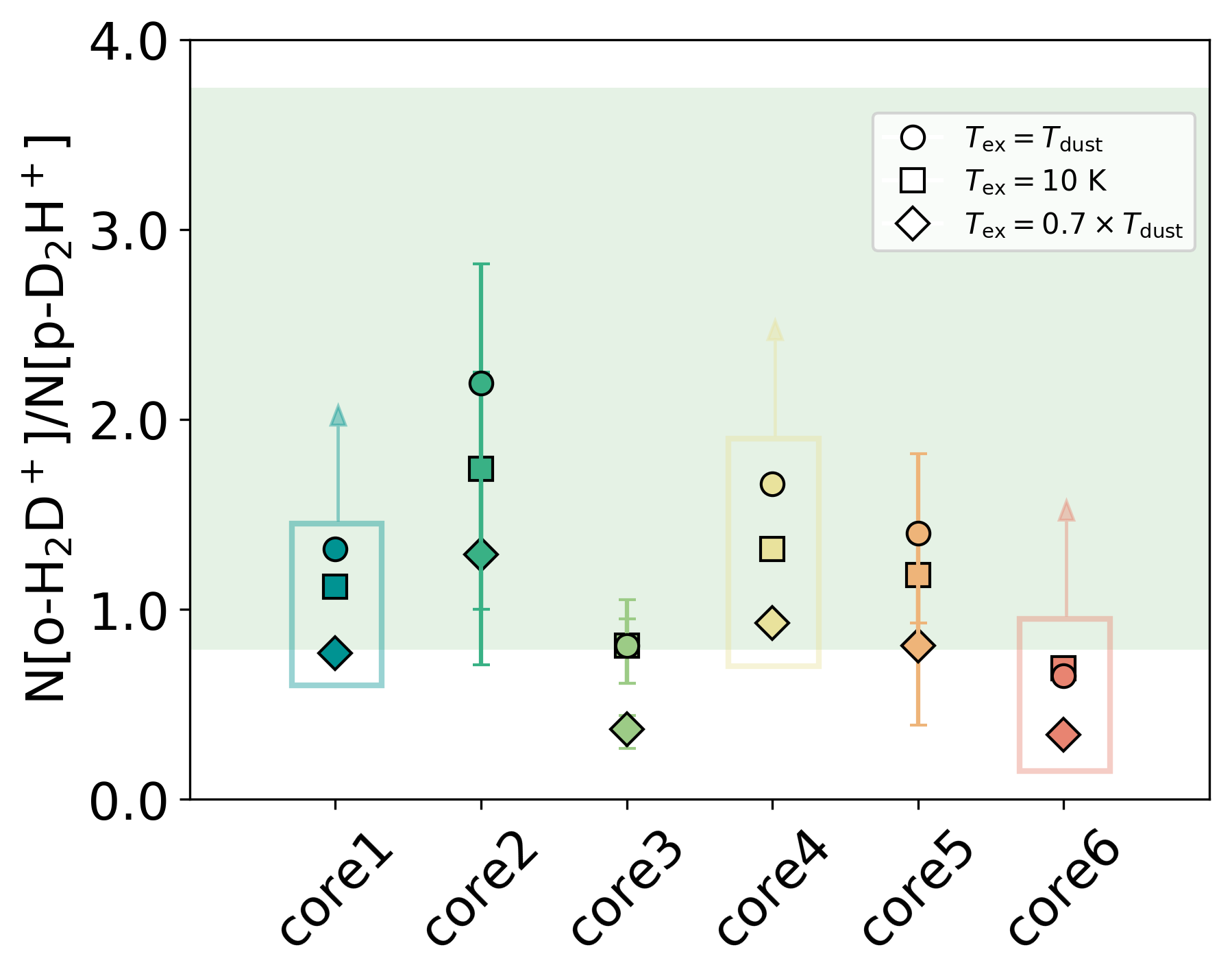}
\caption{N(o-H$_2$D$^+$)/N(p-D$_2$H$^+$) ratio for different $T_\mathrm{ex}$ assumptions. We report three main cases: our reference case $T_\mathrm{ex} = T_\mathrm{dust}$, and two additional cases where we assume lower temperatures, $T_\mathrm{ex} = 10$ K, and $T_\mathrm{ex} = 0.7 \times T_\mathrm{dust}$. The values in the coloured boxes mark the lower limits we calculated for the cases where para-D$_2$H$^+$ was not detected. The green shaded area represents the range of values obtained from our simulations.}\label{fig:tex}
\end{figure}

In the analysis of the observational data, a main assumption has been that $T_\mathrm{ex} = T_\mathrm{dust}$, which may not strictly be the case. A choice for the excitation temperature $T_\mathrm{ex}$ however is needed to fit the spectra and retrieve the column densities of the two deuterated species. To check the dependence on this parameter, we have performed the same fitting procedure outlined in section \ref{sect:analysis}, by assuming $T_\mathrm{ex}=10$ K, and calculated the resulting ortho-to-para ratios. The change in the assumed $T_\mathrm{ex}$ leads to a change of those ratios by roughly 10-15\%. We have also tested an extreme case with 
$T_\mathrm{ex}=0.7\times T_\mathrm{dust}$ following previous works \citep{Caselli2008} and observations of N$_2$H$^+$ of the same region 
 \citep{Friesen2010ApJ}, which report $T_\mathrm{ex}$ slightly lower than those obtained from NH$_3$, with a mean value over the entire Ophiuchus B2 region around 7$\pm 2.8$ K. The differences in this case correspond  to 45\% on the final ortho-to-para ratios. In Fig.~\ref{fig:tex} we report the comparison between the different assumptions for $T_\mathrm{ex}$ (symbols) and the simulations (green shaded area). Overall, if we exclude some outliers with extremely low ortho-to-para ratio (e.g. core 3 and core 6), the assumption for $T_\mathrm{ex}$ does not affect our conclusions on the estimated ages. In the future, the results could be further refined employing a non-LTE analysis, which would however require the observation of an additional line to break degeneracies between free parameters.

\section{Discussion}\label{sect:discussion}
To put our results in the context of the star formation process, we have calculated the ratio between the core age and the dynamical time ($t_\mathrm{age}/t_\mathrm{ff}^0$), where $t_\mathrm{ff}^0$ is the core free-fall time calculated by employing the density of the core at the identification time $n_\mathrm{core}(t_\mathrm{form}$) (see Table \ref{tab:coreprop})\footnote{$n_\mathrm{core}$ is calculated by following an observer-like approach, see also Appendix \ref{sect:appendix}, but we note that there are different approaches. We have tried to calculate the core density as $\Sigma/L$, with $\Sigma$ the surface density and $L$ the depth over which we integrate (i.e. a cylinder of 0.2 pc). This leads to densities on average smaller by a factor of 5-6, which will provide free-fall times a factor of 2 longer, and an overall shorter $t_\mathrm{age}/t_\mathrm{ff}^0$. These numbers are still in line with our main conclusions but we caution that the methodology could affect the results.}. Theories of slow collapse point towards long timescales and strong magnetic or turbulence support, with an initial condition for the cores close to a virial equilibrium where all the forces at play are balanced
 \citep{Shu1987,Mouschovias1991}. Rapid collapse, on the contrary, is based on the idea that the star formation process occurs on typical dynamical timescales \citep{Enrique2005,Hartmann2012} (in this case $t_\mathrm{ff}^0$). All our realizations support the rapid collapse scenario, as the time to reach observed ortho-to-para ratios is of the order of the free-fall time or a few free-fall times (values in between 0.5-5 $t_\mathrm{ff}^0$), and much shorter than the ambipolar diffusion time often invoked by slow-collapse theories \citep{Mouschovias1991}.  The analysis of the ambipolar diffusion time ($t_\mathrm{AD}^0 = 2.5\times 10^{13} x_e$, with $x_e$ the electron fraction,  \citealp[see][]{Shu1987}) in our simulated cores gives values of the order of 400 kyr up to 1 Myr (see Table \ref{tab:coreprop}), which is 4-5 times longer than the estimated cores lifetimes, confirming our conclusions. Our findings are supported by additional observations of the same regions \citep{Pattle2015,Kerr2019}, exploring energy balance and stability, including infall features reported in observed spectra of optically thick tracers  
 \citep{Andre2007,Chitsazzadeh2014,Punanova2016}. These observational studies point towards highly unstable cores with, in some cases, blue-skewed features in the spectra as consequences of infall motions. In addition, estimates in line with our results have been provided from observational statistical methods \citep{Enoch2008}, which by assuming a protostellar stage of 0.54 Myr, obtained an average prestellar lifetime of roughly 500 kyr with an error of a factor of 2.
 
 To check about possible effects on our sources from existing protostars, we reviewed the \textit{Spitzer} results presented in previous works
 \citep{Enoch2009,Dunham2015} and obtained the positions of 23 protostellar candidates over L1688 (18 sources) and L1689 (5 sources). All young stellar objects are located at distances in between 0.02 and 0.12 pc from the cores reported in Table \ref{table:prop}. Among the 18 objects identified, 3 have been classified as Class 0, I protostars, showing an IR-index $\alpha_{IR}\geq 0.3$ \citep{Dunham2015}, 8 as flat-spectrum objects ($-0.3\leq\alpha_{IR}<0.3$) and the remaining 7 as Class II ($-1.6\leq\alpha_{IR}<-0.3$).  Considering their luminosity and applying a simple Stefan-Boltzmann law, we can exclude that radiation feedback can have an effect on the studied cores, consistent with numerical investigations of radiative feedback in low-mass star formation \citep{Offner2009}. 

Two of the sources found in L1688 (J162728.4-242721 and J162730.1-242743) have been associated with an intense proto-stellar activity, identified to potentially be responsible for the single bright and extended gigantic outflow observed at the centre of Oph-B2 \citep{Kamazaki2003,Nakamura2011}. The outflow is extended up to 0.2 pc along the southeast-northwest direction \citep{Kamazaki2019} and has been mapped with several $^{12}$CO, $^{13}$CO and C$^{18}$O low-J molecular transitions. Although encompassing a significant portion of Oph-B2, the outflow does not involve any of the four cores examined in this work. Core 6 appears to be closest to the blueshifted component of the outflow in the northwest direction, but the absence of any relevant broadening in both the observed o-H$_2$D$^+$ and p-D$_2$H$^+$ lines excludes any direct influence of the outflow at least in the region traced by the H$_3^+$ isotopologues. In L1689, three Class 0, I and two Class II protostellar candidates are identified with distances between 0.03 and 0.11 pc around core 1, while no protostellar candidate was found close to core 2. L1689 is also known to be on average much less active than L1688 \citep{Nutter2006}. To our knowledge, there is no evidence for outflows in the immediate surroundings of both of the cores identified in L1689, consistent with the line profiles which indicate no signatures of outflows.

In terms of chemical clocks, previous studies focused on N$_2$D$^+$/N$_2$H$^+$ \citep{Pagani2013}, which is known to have a different chemical evolution compared to the tracers we consider here and to trace mainly later stages of the evolution \citep{Bovino2019,Giannetti2019}. They suggested lifetimes of 500-700 kyr. This is longer than the ages reported in this work but still in line with our current results. 

In addition, we have checked the ratio of the ortho/para H$_2$D$^+$ at the time just before a protostar would form\footnote{Note that the para-H$_2$D$^+$ column density has been convolved with a proper SOFIA beam of \SI{22}{arcsec}.}, and found values as low as 
0.14. This is similar to the ratio of 0.07$\pm$0.03 previously reported \citep{Brunken2014}, but has been obtained in hundreds 
of thousands of years rather than millions of years. The previous measurements \citep{Brunken2014} of course corresponds to the envelope, not to the core, thereby providing an upper limit to the age of the protostar, which is nicely consistent with the observations provided here.  Note that in a subsequent work \citep{Harju2017}, the long timescales previously proposed \citep{Brunken2014} have been lowered to 0.5 Myr as a result of different assumptions and parameter choices in their model. This emphasizes the need to produce self-consistent fully dynamical simulations as pursued here, to obtain the involved timescales from a physical model.

\subsection{Caveats}\label{sect:caveats}
Finally, it is important to notice that, even if our simulations require some assumptions in the initial conditions, these still represent state-of-the art results, due to the computational challenge of including complex chemistry in large scale simulations. So far, chemical clocks have been explored in static/dynamical low-dimensional models, often tuned on observations and evolved on longer timescales, towards equilibrium configurations, without including non-linear effects induced by magnetic fields and turbulence. On the contrary, we start from conditions typical of observed filaments and allow the cores to naturally form in a highly dynamical and complex environment. Some core candidates dissolve during the evolution, some others merge, and in general it takes 50-300 kyr to form the first  cores. This means that starting from an already high density situation is not necessarily speeding-up the formation of the cores and the chemistry associated to them, as non-linear effects induced by turbulence and magnetic fields will act against the formation of overdensities \citep[e.g.][]{Semadeni2005}.

It is also important to note that some of our cores experienced expansion phases before getting into rapid collapse that can be seen as quasi-static stages where the targeted ortho/para ratio is increasing rather than decreasing (see for example F1\_0035\_2 in Fig. \ref{fig:ratio}, or the core evolution reported in Fig. \ref{fig:simmaps2}). This points towards the need of an abrupt change in density to reproduce the observed ratios and then a rapid collapse scenario would be favoured. Even if we cannot exclude that situations of slow-collapse can lead to similar results, we have to consider that previous papers exploring sub-critical filaments \citep[see e.g.][]{Kortgen2018} did not find any fragmentation and no cores formed at all. In addition, none of the cores we formed in our simulations resemble virial conditions at their formation time. This could be the consequence of the idealised setup we are exploring and/or the result of the physics involved in the collapse process. To disentangle between these two scenarios we would need to perform large-scale simulations following the formation of the filaments first and the subsequent fragmentation into small entities, within a highly dynamical environment where accretion processes towards the filaments (that can boost the fragmentation) are also taken into account. Such simulations are currently not available due to the high computational costs, and represent a limitation to a comprehensive understanding of the star-formation process and the usefulness of chemical clocks in disentangling between the different existing star-formation scenarios.

In order to assess how slow-collapse conditions would compare with our three-dimensional framework, in which these situations would have been hard to recreate, we have run a series of single-cell models exploring the speed of the collapse through a parameter $\beta_\mathrm{ff}$ similarly to what has been done for instance in \citet{Kong2015}. In the following, we only summarise the main results, and report all the details in Appendix \ref{appendix2}. Our findings confirm that:
\begin{itemize}
	\item models of slow collapse, say with timescales of 10$\times t_\mathrm{ff}$, are never able to reproduce the physical and chemical properties of the observed cores 
	\item depending on the employed parameters, models with collapse timescales in between $1-5\times t_\mathrm{ff}$ are in line with the observed quantities independently on the parameters, and well agree with our more complex three-dimensional simulations, within the explored parameter space
	\item the ortho-H$_2$D$^+$/para-D$_2$H$^+$ ratio seems to necessarily require fast collapses to reach the observed values and this suggests its ability to distinguish between slow and fast collapse scenarios. In particular slow-collapse would favour lower ortho-to-para ratios (on much longer timescales) because of the long time available to pre-process the chemistry.
\end{itemize}
Overall, the results from the qualitative single-cell models confirm the picture obtained from more complex simulations, and seem to point towards a dynamical scenario driven by a rapid collapse. However, large-scale simulations are needed to completely break the degeneracy and further explore the reliability of this ratio as a chemical clock, together with a statistically relevant observational sample targeting the same tracers.

\section{Conclusions}\label{sect:conclusions}
To conclude, we provide new observational results that suggest that the ortho-H$_2$D$^+$/para-D$_2$H$^+$ ratio might be a reliable chemical clock of star-forming regions. This represents a viable alternative to previously used clocks \citep{Pagani2013,Brunken2014}, due to the possibility to observe the para-D$_2$H$^+$ from the ground, compared to the difficulty to observe para-H$_2$D$^+$ in the THz domain. In addition, we showed that this ratio is a good proxy for the elusive o/p H$_2$ ratio, an aspect that we will explore in a forthcoming paper. Compared to previous works \citep[e.g.][]{Pagani2013}, which ruled out this ratio as chemical clock due to its non-monotonic behaviour, our findings show that this is mainly happening at late stages and is not affecting its effectiveness in tracing the collapse timescale.
We presented a detailed analysis with a comparison of state-of-the-art three-dimensional magneto-hydrodynamical simulations including a detailed chemical model. Our findings are very relevant to understand which physical processes affect the earliest stages of star formation and clarify results derived from more simplified models \citep{Pagani2013,Brunken2014}. 

The comparison between observations and three-dimensional dynamical models provides an overall consistent picture between observations and simulations, where both the chemical abundance ratios as well as dynamical properties such as the virial parameter of the core can be explained with the framework of a rapid collapse. 
The short timescales we have found are, in fact, much shorter than indicated from previous low-dimensional studies
 \citep{Parise2011,Pagani2013,Brunken2014}. 

We aimed to obtain information on the physics of star formation by studying the chemistry of these regions, and the ages are obtained from cutting edge simulations properly compared with observations. From our analysis we conclude that models of rapid collapse are more likely to represent the observed cores, resulting in a prestellar stage of a few hundreds of thousands of years.
However, even if our qualitative single-cell models support our three-dimensional findings and seem to rule out the possibility that slow-collapse could reproduce the observed features, a certain degree of degeneracy should be considered, particularly when taking into account the uncertainties introduced by the assumption of the excitation temperature in the observational analysis. To break this degeneracy it is crucial to perform observations of this chemical clock on a statistical sample of low-mass cores, to exploit the dynamical timescales of regions that could be in different environmental conditions. This will help understanding if rapid collapse is the only possible scenario or if star formation occurs in different modes, depending on the physical and chemical conditions of the clouds.

\section*{Acknowledgements}
The computations/simulations were performed with resources provided by the Kultrun Astronomy Hybrid Cluster at Universidad de Concepci\'on. This paper is based on data acquired with the Atacama Pathfinder EXperiment (APEX). APEX is a collaboration between the Max Planck Institute for Radioastronomy, the European Southern Observatory, and the Onsala Space Observatory.
DRGS aknowledges support from Fondecyt Regular 1201280. AL acknowledges support from MIUR under the grant PRIN 2017-MB8AEZ.




\bibliographystyle{aa}
\bibliography{mybib_D} 



\begin{appendix}
\section{Dynamical quantities}\label{sect:appendix}
The mass and temperature are calculated from the dust continuum spectra, within a circular area of radius $R_\mathrm{eff}$, which represents the area of the extraction region. We employ the \textit{Herschel} \citep{Pilbratt2010} maps at 500, 350 and 250 $\mu$m from SPIRE \citep{Griffin2010} and at 160 $\mu$m and 100 $\mu$m from PACS \citep{Poglitsch2010}, by evaluating the background contribution in two circular regions located in nearby local minima. The H$_2$ total column density peak is calculated following a similar approach. 
The dust continuum spectra is fitted as a grey body with the dust opacity parameters from the ATLASGAL survey
 \citep{Ossenkopf1994,Shuller2009,Konig2017}, while the best fit model is evaluated analogously to previous works \citep{Sabatini2019}. Note that the error on the final observed  mass, H$_2$ column density, and temperature is assumed to be around 20\% \citep{Pagani2015,Schisano2020}.

The Jeans mass is computed as following \citep{Stahler2004,Sadavoy2012}

\begin{equation}\label{eq:jeans}
	M_J = 2.9 \left(\frac{T}{10\, \mathrm{K}}\right)^{1.5} \left(\frac{n_\mathrm{core}}{10^4\, \mathrm{cm^{-3}}}\right)^{-0.5} \,\, M_\odot \,,
\end{equation}

\noindent where the total number density is calculated from the core mass and the volume: $n_\mathrm{core} = \frac{M_{\rm core}}{V_{\rm core}\mu m_\mathrm{H}}$, with $V_{\rm core} = \frac{4}{3}\pi R_{\rm eff}^3$.

Dynamical  quantities are good qualitative indicators of the gas motion. We calculate the virial parameter considering \citep{MacLaren88}

\begin{equation}\label{eq:alpha}
       \alpha = \frac{M_\mathrm{vir}}{M_\mathrm{core}}\:\:\:{\rm with}\:\:\:M_\mathrm{vir} = k_2\:\left(\frac{{R}_\mathrm{eff}}{{\rm pc}}\right)\:\left(\frac{\Delta \upsilon_\mathrm{obs}}{{\rm km\:s}^{-1}}\right)^2\,,
    \end{equation}
\noindent where $k_2=210$, and $\Delta \upsilon_\mathrm{obs}$ is the full-width at half-maximum obtained from the spectra, and $R_\mathrm{eff}$ is reported in Table \ref{table:prop}.

The Mach number, which provides the level of turbulence, is estimated as $\mathcal{M} = \sigma_\mathrm{nth}/c_s$, with the non-thermal velocity dispersion calculated as

\begin{equation}
	\sigma_\mathrm{nth} = \sqrt{\sigma^2_\mathrm{obs} - \sigma^2_\mathrm{th}}
\end{equation}

\noindent where $\sigma_\mathrm{th} = \sqrt{k_b T/m}$ is the thermal velocity dispersion of the molecule, $k_b$ the Boltzmann constant, $m = 6.692 \times 10^{-24}$ the ortho-H$_2$D$^+$ mass in grams, $\sigma_\mathrm{obs} = \Delta \upsilon_\mathrm{obs}/(2\sqrt{2\; \ln 2}$), $c_s = \sqrt{k_b T / (\mu m_\mathrm{H})}$ the thermal sound speed for a particle with mean molecular weight $\mu = 2.33$, and $m_\mathrm{H}$ the hydrogen mass in grams.

We note that, for the sake of consistency in the calculation of the virial parameter for the simulated cores, we added a thermal velocity dispersion component to the gas velocity dispersion obtained from the maps.

\section{Single-cell models}\label{appendix2}
\begin{figure}
	\centering
	\includegraphics[scale=0.45]{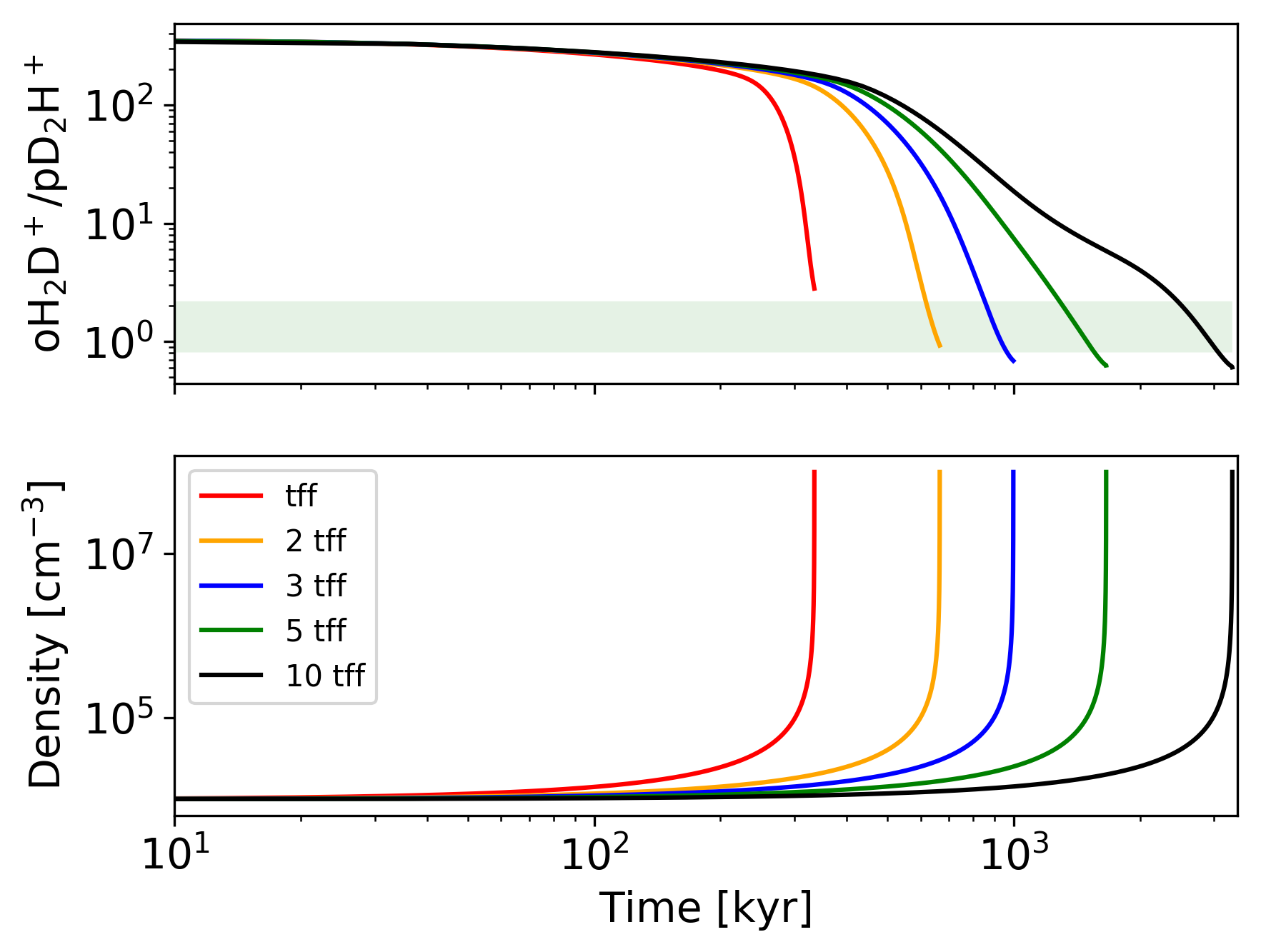}\\
	\includegraphics[scale=0.45]{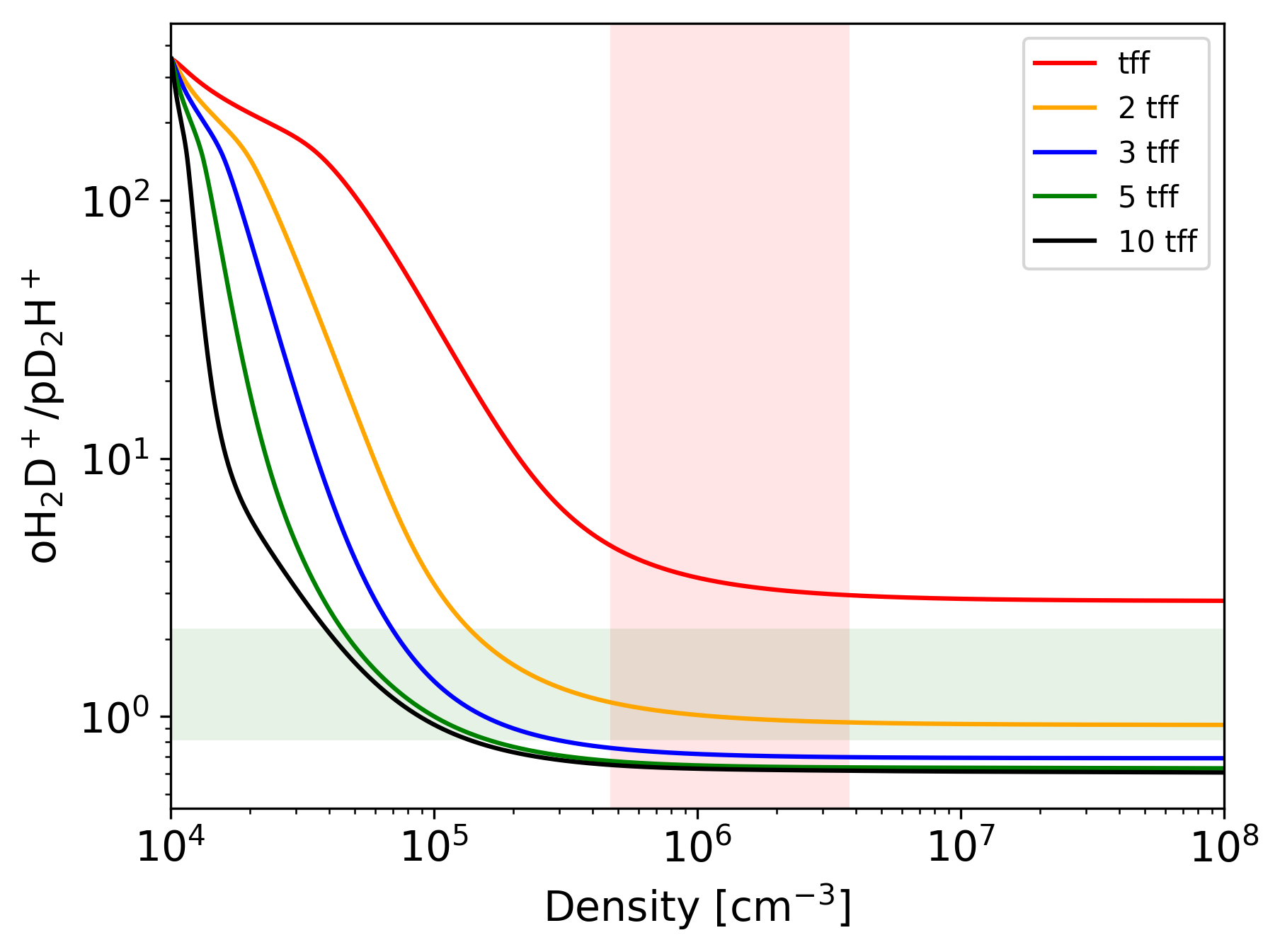}
	\caption{Top: time evolution of the ortho-H$_2$D$^+$/para-D$_2$H$^+$ ratio (top) and density (bottom) for a reference case with $n_\mathrm{core} = 10^{4}$ cm$^{-3}$, $T=10$ K, CRIR = 2.5$\times 10^{-17}$ s$^{-1}$, and initial H$_2$ OPR = 0.1, for different collapse speeds ($\beta_\mathrm{ff}=[1,2,3,5,10]$). The green shaded area represent the range of observed values, only detections are considered in this specific case, see Table \ref{table:results}.  Bottom: Evolution of the ortho-H$_2$D$^+$/para-D$_2$H$^+$ ratio wrt the density for the same parameters. The red shaded are represents the range of densities for the observed cores.}\label{fig:speed}
\end{figure}

To follow the impact of the collapse speed on the final results, and particularly on the timescales to reach observed values, we employed a semi-analytical framework, where the evolution of the density is regulated by the following equation:
\begin{equation}
	\frac{dn_\mathrm{core}(t)}{dt} = \frac{1}{\beta_\mathrm{ff}} \frac{n_\mathrm{core}(t)}{t_\mathrm{ff}}
\end{equation}
\noindent with $t_\mathrm{ff}$ the free-fall time calculated at the density $n_\mathrm{core}(t)$, and $\beta_\mathrm{ff}$ is an arbitrary parameter which determines the collapse speed. We assume $\beta_\mathrm{ff}=[1,2,3,5,10]$, where the extreme cases represent: i) a collapse occurring on the free-fall timescale ($\beta_\mathrm{ff}=1$), and ii) a collapse on the ambipolar diffusion time (i.e. on a timescale of 10$\times t_\mathrm{ff}$ with $\beta_\mathrm{ff} = 10$, \citealp[see e.g.][]{Semadeni2005}). We have explored a large parameter space, by changing the initial ortho-to-para H$_2$ ratio (0.1 and 10$^{-3}$), the initial collapse density ($n_\mathrm{core} =$ 10$^{3}$,10$^4$, and 10$^5$ cm$^{-3}$), the cosmic ray ionization rate $\zeta_2$ ($1.3\times 10^{-17}$ s$^{-1}$, $2.5\times 10^{-17}$  s$^{-1}$, and $5.0\times10^{-17}$ s$^{-1}$), and the temperature ($T=10$ and 15 K). The initialization of the chemical species is identical to the one used in the three-dimensional simulations. The main intent of these single-cell models is to mimic different dynamical environments, particularly in terms of collapse speed, and see under which conditions we can reach the observed densities and oH$_2$D$^+$/pD$_2$H$^+$ ratios. The list of performed runs is reported in Table \ref{table:singlecells}.

\begin{table*}
\caption{Details of the performed single-cell models. We report the dust temperature, the cosmic ray ionization rate $\zeta_2$, the initial H$_2$ ortho-to-para ratio, the initial density $n_\mathrm{core}$, and the collapse speed parameter $\beta_\mathrm{ff}$. Note that $T = T_\mathrm{dust} = T_\mathrm{gas}$.}\label{table:singlecells}
\centering
\begin{tabular*}{\textwidth}{@{\extracolsep{\fill} } lccccc}
\hline 
\hline
\# Run & $T$ & $\zeta_2$ &  OPR(H$_2$)  &  $n_\mathrm{core}$ & $\beta_\mathrm{ff}$\\
 & [K] & [$10^{-17}$ s$^{-1}]$& & [cm$^{-3}$] &  \\
 \hline
A &  15 & 2.5  & 0.1 & 10$^{4}$ & [1,2,3,5,10] \\
B &  10 & 2.5  & 0.1 & 10$^{4}$ & [1,2,3,5,10]\\
C &  15 & 1.3  & 0.1 & 10$^{4}$ & [1,2,3,5,10]\\
D &  10 & 1.3  & 0.1 & 10$^{4}$ & [1,2,3,5,10]\\
E &  15 & 5.0  & 0.1 & 10$^{4}$ & [1,2,3,5,10]\\
F &  10 & 5.0  & 0.1 & 10$^{4}$ & [1,2,3,5,10]\\
G &  15 & 2.5  & 0.1 & 10$^{3}$ & [1,2,3,5,10]\\
H &  10 & 2.5  & 0.1 & 10$^{3}$ & [1,2,3,5,10]\\
I  &  15  & 2.5  & 0.1 & 10$^{5}$ & [1,2,3,5,10]\\
J &  10  & 2.5  & 0.1 & 10$^{5}$ & [1,2,3,5,10]\\
K &  10  & 2.5  & 10$^{-3}$ & 10$^{4}$ & [1,2,3,5,10]\\
\hline
\hline
\end{tabular*}
\end{table*}

We assume as a reference case the one with initial density $n_\mathrm{core} = 10^4$ cm$^{-3}$, a temperature of $T = 10$ K, and $\zeta_2 = 2.5\times 10^{-17}$ s$^{-1}$, and an initial \mbox{ortho-/para- H$_2$ = 0.1.} The results for this reference case are reported in Fig. \ref{fig:speed}. We show the oH$_2$D$^+$/pD$_2$H$^+$ ratio (OPR) and the density evolution over time, for the different collapse cases. We can see how the collapse is delayed when we increase the $\beta_\mathrm{ff}$ parameter, going from a collapse occurring on the $t_\mathrm{ff}\sim 330$ kyr (red lines), to one occurring on a scale of a few Myr ($10\times t_\mathrm{ff}$), which should resemble the ambipolar diffusion time (black lines). Rapid collapse cases reach a low OPR in very short times, mainly driven by the changes in density, while the slow collapse case evolves at roughly constant density for at least one Myr, allowing for a pre-processing of the chemistry, which leads to a low OPR even before the onset of the rapid contraction. Overall when the collapse kicks in, the OPR values are lower than in the fast collapse cases. This can be seen also by looking at the bottom panel of Fig. \ref{fig:speed}, where we show the evolution of the OPR with density. All the cases reach an equilibrium OPR at densities similar to the observed ones (red shaded region). By comparing the obtained OPR with the observed ones (green shaded area) we can see that very slow collapses (i.e. 5-10 $t_\mathrm{ff}$), lie at the lower limit of the observed values, while collapses on the order of 2-3$\times t_\mathrm{ff}$, representing fast collapses, are in line with observations. 

\begin{figure}
	\centering
	\includegraphics[scale=0.45]{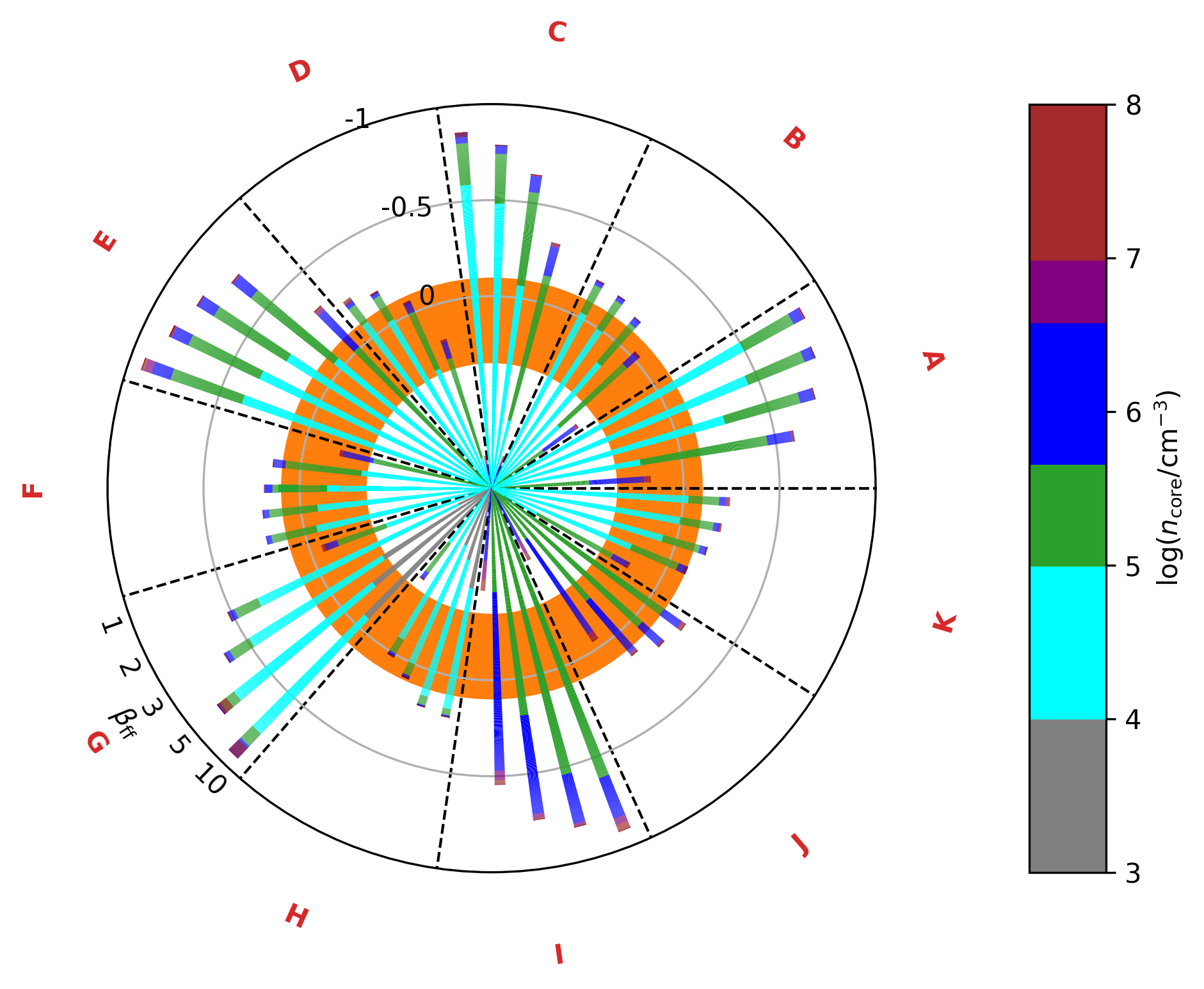}
	\caption{Polar-like chart summarising the results of our parameter study. Each slice represents a model (labeled with letters) with fixed initial conditions in terms of $T$, $\zeta_2$, initial density $n_\mathrm{core}$, and initial H$_2$ OPR (see Table \ref{table:singlecells}).  Within the model slice the different collapse speed cases defined by the parameter $\beta_\mathrm{ff}$ are represented by a stacked bar. The radial evolution of the bar reflects the oH$_2$D$^+$/pD$_2$H$^+$ evolution and the bars are colored by density (see associated colorbar). The orange ring is delimited by the observed oH$_2$D$^+$/pD$_2$H$^+$ ratio and the blue region in the colorbar is set by the density of the observed cores (see also Table \ref{table:prop}). This means that we have a match between the models and the observations when the blue frame of the bar falls into the orange region.}\label{fig:parameter}
\end{figure}

To have a broader overview of the effects of the collapse speed on the final OPR, we show the results of the large parameter study in Fig. \ref{fig:parameter}, in which the different sectors (enclosed within the black dashed lines and associated to the letters from A to K) correspond to the models in Table \ref{table:singlecells}, for which the five values of $\beta_\mathrm{ff}$ are reported counterclock-wise. The orange ring in the polar-chart like plot represents the observed values, and the blue color in the stacked bars reflects the observed densities. A match between models and observations is in place when the blue portion of the stacked bar falls in the orange region.
We can appreciate that the slow-collapse cases are always at the limit of the observed OPR values, or outside the range, because the long-time available to pre-process the chemistry allows to reach lower OPR at the observed densities. This is particular true for the case with 10$\times t_\mathrm{ff}$ that never enters the observed region defined by the orange ring and the blue frame of the bar. On average, the wide range of parameters explored points towards a good matching for the cases with collapse times of 1-3$\times t_\mathrm{ff}$, in agreement with our more complex three-dimensional simulations. There are a few cases where also 5$\times t_\mathrm{ff}$ is falling in the observed region when $T=10$ K and the initial density is 10$^5$ cm$^{-3}$, with a CRIR=2.5$\times$10$^{-17}$ s$^{-1}$ and an initial H$_2$ OPR of 0.1. Collapses on the free-fall timescale are also giving good results but show some discrepancies when the temperature is 10 K and the initial CRIR is low, i.e. 1.3$\times 10^{-17}$ s$^{-1}$.
Even if this is a somewhat qualitative framework, it confirms that slow-collapse cases are not in line with observations and in general reach lower OPR at lower densities. The evolution of deuterated species is indeed driven both by time and density, leading to degeneracy, depending if there is long-time to process the chemistry at almost constant density, or if there is an abrupt change of density in short time. Chemistry alone can lead to degeneracy in the results and in their interpretation. It is then fundamental when comparing with the observations to match also the physical and dynamical quantities and not only the chemistry. Our 3D simulations (see Fig. \ref{fig:simcomp}), not only reproduce the chemical features but also the dynamical and physical properties of the cores, and this together with the results of the qualitative single-cell models, points towards a rapid collapse scenario for the six cores observed in Ophiuchus. As a final note, the current comparison has been performed with our fiducial observational results (i.e. $T_\mathrm{ex} = T_\mathrm{dust}$) and only considering the detections. If we widen the analysis by including the uncertainties induced by the $T_\mathrm{ex}$ assumption and by considering also the lower-limits, we will have a few cases where also slow-collapse will match the observations. However, due to the large number of uncertainties, we can conclude that the majority of the single-cell models point towards fast rather than slow-collapses.

\end{appendix}
\end{document}